\documentstyle[psfig]{mn}

\def\etal{{\it et al.\ }}
\def\eg{{\it e.g.\ }}

\def\spose#1{\hbox to 0pt{#1\hss}}
\def\approxlt{\mathrel{\spose{\lower 3pt\hbox{$\sim$}}
	\raise 2.0pt\hbox{$<$}}}
\def\approxgt{\mathrel{\spose{\lower 3pt\hbox{$\sim$}}
	\raise 2.0pt\hbox{$>$}}}
\def\approxpropto{\mathrel{\spose{\lower 3pt\hbox{$\sim$}}
	\raise 2.0pt\hbox{$\propto$}}}
\mathchardef\twiddle="2218

\def\multleft#1{\hbox to size{\vbox {\halign {\lft{##}\cr #1}}\hfill}\par}
\def\multright#1{\hbox to size{\vbox {\halign {\rt{##}\cr #1}}\hfill}\par}
\def\Mdot{\hbox{$\dot M$}}

\def\today{\ifcase\month\or January\or February\or March\or April\or May\or
      June\or July\or August\or September\or October\or November\or December\fi
      \space\number\day, \number\year}
\def\<{\thinspace}

\def\apc{\rm atom cm$^{-2}$}

\def\cm{{\rm\thinspace cm}}
\def\erg{{\rm\thinspace erg}}

\def\K{{\rm\thinspace K}}
\def\keV{{\rm\thinspace keV}}

\def\km{{\rm\thinspace km}}

\def\Mpc{{\rm\thinspace Mpc}}
\def\Msun{\hbox{$\rm\thinspace M_{\odot}$}}

\def\s{{\rm\thinspace s}}
\def\yr{{\rm\thinspace yr}}

\def\ergpcmsqps{\hbox{$\erg\cm^{-2}\s^{-1}\,$}}

\def\ergps{\hbox{$\erg\s^{-1}\,$}}

\def\kmps{\hbox{$\km\s^{-1}\,$}}

\def\kmpspMpc{\hbox{$\kmps\Mpc^{-1}$}}

\def\apc{\rm atom cm$^{-2}$}
\title{Hard X-ray emission from elliptical 
galaxies} 
\author[Hard X-ray emission from elliptical galaxies]
{\parbox[]{6.in} {S.W. Allen$^1$, T. Di Matteo$^2\thanks{Chandra Fellow}$ and 
A.C. Fabian$^1$ \\
\footnotesize
1. Institute of Astronomy, Madingley Road, Cambridge CB3 0HA\\
2. Harvard-Smithsonian Center for Astrophysics, 60 Garden Street, Cambridge MA 02138, USA
}}
\begin{document}
\date{}
\maketitle

\begin{abstract}
\noindent We report the detection of hard X-ray emission
components in the spectra of six nearby, giant elliptical
galaxies observed with the ASCA satellite. The systems studied, which
exhibit strong dynamical evidence for supermassive black holes in
their nuclei, are M87, NGC 1399 and NGC 4696 (the dominant galaxies of
the Virgo, Fornax and Centaurus clusters, respectively) and NGC 4472,
4636 and 4649 (three further giant ellipticals in the Virgo cluster).
The ASCA data for all six sources provide clear evidence for hard
emission components, which can be parameterized by power-law models with 
photon indices in the range $\Gamma = 0.6-1.5$ (mean value 1.2) and 
intrinsic $1-10$ keV luminosities of $2 \times 10^{40}-2 \times 10^{42}$ 
\ergps. Our results imply the identification of a new class of accreting X-ray
source, with X-ray spectra significantly harder than those of binary
X-ray sources, Seyfert nuclei or low luminosity AGN, and bolometric
luminosities relatively dominated by their X-ray emission. We discuss
various possible origins for the hard X-ray emission and argue that it
is most likely to be due to accretion onto the central supermassive
black holes, via low-radiative efficiency accretion coupled with
strong outflows. In the case of M87, our detected power-law flux is in
good agreement with a previously-reported measurement from ROSAT High
Resolution Imager observations, which were able to resolve the jet
from the nuclear X-ray emission components. We confirm previous
results showing that the use of multiphase models in the analysis of
the ASCA data leads to determinations of approximately solar
emission-weighted metallicities for the X-ray gas in the galaxies. We
also present results on the individual element abundances in NGC 4636.

\end{abstract}

\section{Introduction}

Observations of nearby galaxies provide convincing evidence for the
existence of supermassive black holes.  Although most such galaxies
exhibit little or no nuclear activity, dynamical arguments based on
the observed stellar and gas distributions firmly imply the presence
of supermassive compact objects in their cores (\eg Kormendy \&
Richstone 1995, Magorrian \etal 1998, Ho 1998, van der Marel 1999). 
Interestingly, these studies show that virtually all early-type
galaxies host black holes with masses in the range $10^8-$ a few $10^9$
\Msun. In disk galaxies, however, the (rather sparser) observations
indicate that the black hole masses are typically of order $10^7 \Msun$ (\eg
SgrA$^*$ in our own Galaxy, M31, and the maser galaxy NGC 4258),
consistent with the black hole masses in galaxies being roughly
proportional to the masses of the spheroidal components.

Although the central black hole masses in early type galaxies are
large enough to be the remnants of QSO phenomena (with luminosities
ranging from $10^{46}$ to $10^{48} \ergps$) and giant ellipticals are
known to host radio galaxies and radio-loud quasars at high redshifts,
in nearby ellipticals, only low-luminosity radio cores are commonly
observed (Sadler, Jenkins \& Kotanji 1989; Wrobel \& Heershen
1991). Spiral galaxies, in contrast, frequently exhibit nuclear
activity (\eg Seyfert nuclei), with optical/UV emission, strong,
power-law X-ray continua (with a typical intrinsic photon index,
$\Gamma \sim 2$) and complex variability observed. Such phenomena are
usually attributed to standard, thin-disk accretion, coupled with a
hot coronal plasma with high radiative efficiencies.

Postulating that the early-type systems are simply `starved' of fuel
to power their activity appears implausible since these galaxies
possess extensive hot, gaseous halos which will inevitably accrete
onto their central black holes at rates  which may be estimated
using the Bondi (1952) formula (a lower limit). This accretion should
give rise to far more activity than is observed, if the radiative
efficiency is $\sim 10$ per cent, as is postulated in standard
accretion theory (\eg Fabian \& Canizares 1988). All nearby elliptical
galaxies should then host active nuclei with luminosities exceeding
$10^{45} \ergps$. However, the available data show that the {\it
total} X-ray luminosities of these galaxies rarely exceed $10^{42}
\ergps$, with only a small fraction of this flux being due to a
central point source.

Accretion with such high radiative efficiency need not be
universal, however. As suggested by several authors (Rees \etal 1982; 
Begelman, Blandford \& Rees 1984; Fabian \& Rees 1995) and
successfully applied to a number of giant ellipticals in the Virgo
cluster (Reynolds \etal 1996; Di Matteo \& Fabian 1997a; Mahadevan
1997; Di Matteo \etal 1999a), the final stages of accretion in
elliptical galaxies may occur via an advection-dominated accretion
flow (ADAF; Narayan \& Yi 1995, Abramowicz \etal 1995) at roughly the
Bondi rates. Within the context of such an accretion mode, the
quiescence of the nuclei in these systems is not surprising; when the
accretion rate is low, the radiative efficiency of the accreting (low
density) material will also be low. Other factors may also contribute
to the low luminosities observed. As discussed by Blandford \&
Begelman (1999; and emphasized observationally by Di Matteo \etal
1999a), winds may transport energy, angular momentum and mass out of
the accretion flows, resulting in only a small fraction of the
material supplied at large radii actually accreting onto the central
black holes.

If the accretion from the hot interstellar medium in elliptical
galaxies (which should have relatively low angular momentum) proceeds
directly into the hot, advection-dominated regime, and low-efficiency
accretion is coupled with outflows (Di Matteo \etal 1999a), the
question arises of whether {\it any} of the material entering into the
accretion flows at large radii actually reaches the central black holes.  
The present observational data generally provide little or no evidence for
detectable optical, UV or X-ray emission associated with the nuclear
regions of these galaxies.

In this paper we present a detailed, multiphase analysis of high-quality 
ASCA X-ray spectra for six nearby elliptical galaxies; three central 
cluster galaxies and three giant ellipticals in the Virgo 
Cluster. We obtain clear detections of hard (mean-weighted photon index, 
$\Gamma=1.2$) power-law, emission components in the integrated X-ray 
spectra of all six galaxies. Although our data do not allow us to 
unambiguously identify these components with the nuclear regions of the 
galaxies, we show that this emission is likely to be related 
to the accretion process and the presence of the central, supermassive 
black holes, which also power radio activity in the galaxies at varying 
levels.

The presence of hard X-ray emission components in the ASCA 
spectra of elliptical galaxies has previously been reported by a number
of authors (\eg Matsushita \etal 1994; Matsumoto \etal 1997; Buote 
\& Fabian 1998). In general, however, these studies have associated the 
observed, hard components with a population of binary X-ray sources, the 
spectra of which can be approximated by bremsstrahlung models with typical 
temperatures of $4-7$ keV (\eg Fabbiano, Trinchieri \& Van Speybroeck 1987; 
White, Stella \& Parmar 
1988; Makishima \etal 1989; Tanaka \etal 1996; Christian \& Swank 1998).
The spectra of the hard X-ray emission components detected in the present 
sample of elliptical galaxies are significantly harder than those of binary 
X-ray sources or typical AGN, identifying these objects as, potentially, a 
new class of accreting source. (The `canonical' photon index for AGN has  
$\Gamma \sim 1.8-2$; Nandra \etal 1997, Reynolds 1997; recent ASCA 
studies of low-luminosity AGN indicate photon indices in the range $\Gamma 
\sim 1.6-1.8$; \eg M81; Ishiaki \etal 1996, Turner \etal 1996 or NGC 4258; 
Makishima \etal 1994) The X-ray luminosities of the present  systems 
($L_{\rm X,1-10} = 2 \times 10^{40}- 2 \times 10^{42}$ \ergps) are similar to 
those of low  luminosity AGN (\eg Ho, Filippenko \& Sargent 1997), although 
the large ($\sim 10^{9}$ \Msun) black hole masses in their nuclei identifies 
them  as far more inefficient radiators. The spectral energy distributions for 
the galaxies in our sample indicate that a significant fraction of their 
luminosities are emitted at X-ray
wavelengths, with relatively low levels of optical and UV emission,
which are dominant in typical AGN. These data, coupled with the
presence of compact (possibly thermal synchrotron) radio cores,
suggest a possible, ubiquitous presence of low-level nuclear activity
in the nearby universe and provide important new constraints on the
dominant accretion mechanisms in elliptical galaxy cores.

\section{Observations and data reduction}

\begin{table*}
\begin{center}
\caption{Summary of the ASCA Observations}
\vskip 0.1truein
\begin{tabular}{ c c c c c c c c c }
 \hline                                                                               
Cluster         & ~ &  $D_{\rm L}$ &  Date  & ~ &    S0  &   S1  &  G2   &  G3  \\  
 \hline                                                                               
&&&&&&&& \\                                                                         
M87             & ~ & 18.0 &  1993 Jun 07  & ~ &  13633 & 14668 & 16874 & 16874 \\      
NGC 1399(1)     & ~ & 29.0 &  1993 Jul 15  & ~ &  16636 & 17678 & 19646 & 19646 \\       
NGC 1399(2)     & ~ & ---  &  1993 Jul 16  & ~ &  14266 & 16961 & 19741 & 19785 \\       
NGC 4696        & ~ & 62.6 &  1995 May 19  & ~ &   ---  & 67650 & 70981 & --- \\       
NGC 4472(1)     & ~ & 18.0 &  1993 Jun 30  & ~ &  17022 & 18314 & 22242 & 22240 \\       
NGC 4472(2)     & ~ & ---  &  1993 Jul 04  & ~ &  16461 & 18781 & 22739 & 22731 \\       
NGC 4636        & ~ & 18.0 &  1995 Dec 28  & ~ &  244180 & 245450 & 183290 & 183280 \\       
NGC 4649(1)     & ~ & 18.0 &  1994 Jan 07  & ~ &  21438 & 10669 & 40928 & 40926 \\       
NGC 4649(2)     & ~ & 18.0 &  1994 Jan 07  & ~ &  13775 & ---   & ---   & ---  \\       

&&&&&&&& \\                                                                         
\hline 
&&&&&&&& \\                                                                         
\end{tabular}
\end{center}
\parbox {7in} {Notes: Column 2 lists the luminosity distances ($D_{\rm
L}$) to the galaxies in Mpc. For the Virgo Cluster galaxies we assume
a fixed distance of 18Mpc. For NGC 1399 and 4696 the
distances are calculated using the optically-determined
redshifts and an assumed cosmology of $H_0$=50 \kmpspMpc, $\Omega = 1$
and $\Lambda = 0$.  Column 3 lists the dates of the
observations. Columns $4-7$ summarize the net exposure times (in
seconds) in the four ASCA detectors, after all screening and cleaning
procedures were carried out.  For NGC 4696 only data from
the S1 and G2 detectors were included in the analysis, due to residual
calibration errors in the data from the other detectors for that
observation. For NGC 1399, 4472 and 4649, the observations were made
in two parts.  For the second NGC 4649 observation, only the S0 data
were of sufficient quality to be of use in the analysis.}

\end{table*}

\subsection{Target selection}

The targets selected for investigation are six well-studied, nearby
elliptical galaxies, with high-quality ASCA observations available in
public archives. The objects include three dominant cluster galaxies;
the well known, low-luminosity radio galaxy M87 (NGC 4486) at the
centre of the Virgo Cluster; NGC 1399, the giant elliptical at the
centre of the Fornax Cluster and NGC 4696, the dominant galaxy of the
Centaurus Cluster. These three galaxies, and especially M87, are key
examples of quiescent active nuclei, containing nuclear black hole
masses of $3-5 \times 10^9 \Msun$ (Ford \etal 1995; Harms \etal 1994;
Macchetto \etal 1998; Magorrian \etal 1998; although no direct mass
measurement for NGC 4696 has been made) and exhibiting low-luminosity
relativistic jets and FR-I-type radio emission. 
However, the luminosities of their nuclei are approximately
three orders of magnitude less than would be expected if they were
accreting at close to their Bondi rates, with a standard radiative
efficiency (\eg Reynolds \etal 1996).

The sample also includes three other giant ellipticals in the Virgo
cluster; NGC 4649, 4472 and 4636. These galaxies contain almost
completely inactive black holes, with measured masses ranging from a
few $10^8 \Msun$ to a few $10^9$\Msun (Magorrian \etal 1998) and
predicted X-ray luminosities (if the accretion occurs at the Bondi
rate with a standard radiative efficiency) four to five orders of
magnitude greater than is observed (Di Matteo \etal 1999a).

The central cluster galaxies provide an extreme illustration of the
phenomenon of quiescent black holes. Beside possessing FR-I-type radio
sources and very large black hole masses, these galaxies exist in
extremely gas-rich environments {\it i.e.} in cooling flows at the
centres of clusters, and are therefore ideal sources in which to study
the physics of low-radiative efficiency accretion. However, these same
reasons also imply that these galaxies are by no means
typical. Despite exhibiting lower luminosities, the properties of the
other three ellipticals in our sample, which do not exist in such
preferential locations, may be more easily generalized to other
systems. These three galaxies have also been studied at high radio
frequencies and sub-mm wavelengths (Di Matteo \etal 1999a) and, when
coupled with the reduced jet-radio-activity in these systems (and the
ASCA data presented in this paper), provide further crucial
constraints on the primary accretion mechanisms in the nuclei of
elliptical galaxies.

\subsection{The ASCA Observations and data reduction}

The ASCA (Tanaka, Inoue \& Holt 1994) observations were made over a
two-and-a-half year period between 1993 June and 1995 December.  The
ASCA X-ray telescope array (XRT) consists of four nested-foil
telescopes, each focussed onto one of four detectors; two X-ray CCD
cameras, the Solid-state Imaging Spectrometers (SIS; S0 and S1) and
two Gas scintillation Imaging Spectrometers (GIS; G2 and G3). The XRT
provides a spatial resolution of $\sim 3$ arcmin (Half Power Diameter)
in the energy range $0.5 - 10$ keV. The SIS detectors provide
excellent spectral resolution [$\Delta E/E = 0.02(E/5.9 {\rm
keV})^{-0.5}$] over a $22 \times 22$ arcmin$^2$ field of view. The GIS
detectors provide poorer energy resolution [$\Delta E/E = 0.08(E/5.9
{\rm keV})^{-0.5}$] but cover a larger circular field of view of $\sim
50$ arcmin diameter.

For our analysis of the ASCA data we have used the screened event
lists from the rev2 processing of the data sets available on the
Goddard-Space Flight Center (GSFC) ASCA archive (for details see the
ASCA Data Reduction Guide, published by GSFC.) The data were reduced
using the FTOOLS software (version 4.1) from within the XSELECT
environment (version 1.4).  Further data-cleaning procedures as
recommended in the ASCA Data Reduction Guide, including appropriate
grade selection, gain corrections and manual screening based on the
individual instrument light curves, were followed.  A full summary of
the observations is given in Table 1.

Spectra were extracted from all four ASCA detectors, except for NGC
4696 for which the S0 and G3 data exhibited residual calibration
errors and so were excluded from the analysis. The spectra were
extracted from circular regions, centred on the peaks of the X-ray
emission from the galaxies. For the SIS data, the radii of the regions
studied were selected to minimize the number of chip boundaries
crossed (thereby minimizing the systematic uncertainties introduced by
such crossings) whilst covering as large a region of the galaxies as
possible. Data from the regions between the chips were masked out and
excluded. The final extraction radii for the SIS data are summarized
in Table 2. Also noted are the chip modes used for the observations
and the number of chips from which the extracted data were drawn.  For
the GIS data, a fixed extraction radius of 6 arcmin was adopted.

Background subtraction was carried out using the `blank sky'
observations of high Galactic latitude fields compiled during the
performance verification stage of the ASCA mission. The background
spectra were screened and grade selected in the same manner as the
target observations and extracted from the same regions of the
detectors. For the systems observed in 4-CCD mode, additional SIS
background spectra were extracted from regions of the chips free from
significant source counts (we note that this was not possible for the
M87 observation due to the extended cluster emission which covers the
full area of the detectors). The use of these on-chip backgrounds in
place of the blank-sky backgrounds did not significantly alter the
results.

For the SIS data, response matrices were generated using the FTOOLS
SISRMG software (version 1.1). Where the extracted spectra covered
more than a single chip, individual response matrices were created for
each chip, which were then combined to form a counts-weighted mean
matrix. For the GIS analysis, the response matrices issued by GSFC on
1995 March 6 were used.

\begin{table}
\begin{center}
\caption{Extraction radii and chip modes for the SIS observations}
\vskip 0.1truein
\begin{tabular}{ c c c c c }
\hline                                                          
Cluster         & ~ & S0           & S1             & Chip Mode \\  
                & ~ & (amin/kpc) & (amin/kpc)   &           \\  
\hline                                                         
&&&& \\                                                   
M87             & ~ & 4.9/25.5  & 4.1/21.3    & 4(2)    \\      
NGC 1399(1)     & ~ & 4.2/35.1  & 3.7/30.9    & 4(1)    \\      
NGC 1399(2)     & ~ & 4.3/35.9  & 4.2/35.1    & 4(1)    \\      
NGC 4696        & ~ & ---       & 3.7/66.0    & 1(1)    \\      
NGC 4472(1)     & ~ & 3.7/19.3  & 3.7/19.3    & 4(1)    \\      
NGC 4472(2)     & ~ & 3.2/16.7  & 3.4/17.7    & 4(1)    \\      
NGC 4636        & ~ & 4.4/22.9  & 3.6/18.7    & 1(1)    \\      
NGC 4649(1)     & ~ & 3.8/19.8  & 2.9/15.1    & 4(2)    \\      
NGC 4649(2)     & ~ & 3.8/19.8  & ---         & 2(2)    \\      

&&&& \\                                                                     
\hline 
&&&& \\                                                                     
\end{tabular}
\end{center}
\parbox {3.5in}
{The radii of the circular extraction regions for the SIS spectra (in
arcmin and kpc) and the chip modes used in the observations (either 1,2 or 4-CCD mode). 
The numbers in parentheses indicate the number of 
chips contributing to the extracted spectra. For the GIS data, a fixed
extraction radius of 6 arcmin was used.}
\end{table}

\section{The Analysis of the ASCA data}

\subsection{The spectral models} 

The modeling of the X-ray spectra has been carried out using the XSPEC
spectral fitting package (version 10.0; Arnaud 1996). For the SIS
data, only counts in pulse height analyser (PHA) channels
corresponding to energies between 0.6 and 10.0 \keV~ were included in
the analysis (the energy range over which the calibration of the SIS
instruments is best-understood). For the GIS data, only counts in the
energy range $1.0 - 10.0$ \keV~were used. The spectra were grouped
before fitting to ensure that $\chi^2$ statistics could be reliably
used (after background subtraction).

The spectra have been modeled using the plasma codes of Kaastra \&
Mewe (1993; incorporating the Fe L calculations of Liedahl, Osterheld
\& Goldstein 1995) and the photoelectric absorption models of
Balucinska-Church \& McCammon (1992). The data from all four detectors
were included in the analysis, with the fit parameters linked to take
the same values across the data sets. The exceptions to this were the
emission measures of the (hot) plasma components, which model the
extended X-ray halos of the galaxies and which, due to the different
extraction radii used for the different detectors, were maintained as
independent fit parameters.

The spectra were examined with a series of spectral models. (We have
adopted the naming convention of Allen \etal 1999 from their analysis
of nearby cluster spectra). The first model, Model B, consists of an
isothermal plasma of temperature, $kT$, and metallicity, $Z$, in
collisional equilibrium, at the optically-determined redshift for the
galaxy, and absorbed by a column density $N_{\rm H}$. Metallicities
are measured relative to solar photospheric values of Anders \&
Grevesse (1989) with the various elements assumed to be present in
their solar ratios. The second model, model D, included an additional,
cooler plasma component of temperature, $kT_2$, the normalization of
which was a further free parameter in the fits. (This cooler
component, where present, is typically associated with the presence of
a cooling flow in a cluster or galaxy).  The metallicity of the cooler
component was linked to that of the hotter gas. The cooler component
was also assumed to be absorbed by an intrinsic column density,
$\Delta N_{\rm H}$, of cold gas, which was a further free parameter in
the fits. The abundances of metals in the absorbing gas were fixed at
their solar values (Anders \& Grevesse 1989).

Where it provided a significant statistical improvement, a third
spectral model, model F, was also examined which was similar to model
D but with the abundances of various individual elements also included
as free parameters in the fits. This model was statistically preferred
over model D for M87, NGC 4696 and NGC 4636.

Finally, a fourth spectral model, Model G, was investigated in which a
power-law emission component was introduced into the previous
best-fitting two-temperature plasma model (either model D or F,
respectively). The power-law emission component was assumed to be
absorbed by the same intrinsic column density acting on the cooler
plasma emission component (since both components are expected to arise
primarily from the central regions of the galaxies and/or clusters).

\subsection{The requirement for multi-phase models and the metallicity 
of the X-ray gas}

\begin{table*}
\begin{center}
\caption{Spectral results for the central cluster galaxies}
\vskip 0.1truein
\begin{tabular}{ c c c c c c c c c c c }
\hline 
& ~ & Parameters & ~ & Model B & ~~~ & Model D & ~~~ & Model F & ~~~ & Model G \\ 
\hline 
&&&&&&&&&& \\ 
& ~ & $kT$ & ~ &
$2.02^{+0.01}_{-0.02}$ & ~~~ & $2.19^{+0.02}_{-0.03}$ & ~~~ &
$2.17^{+0.03}_{-0.03}$ & ~~~ & $2.01^{+0.04}_{-0.03}$ \\ & ~ & $Z_{\rm
Fe}$ & ~ & $0.65^{+0.02}_{-0.02}$ & ~~~ & $0.79^{+0.04}_{-0.03}$ & ~~~
& $0.76^{+0.04}_{-0.03}$ & ~~~ & $0.72^{+0.04}_{-0.04}$ \\ M87 & ~ &
$N_{\rm H}$ & ~ & $0.35^{+0.03}_{-0.04}$ & ~~~ &
$0.31^{+0.07}_{-0.07}$ & ~~~ & $0.27^{+0.07}_{-0.07}$ & ~~~ &
$0.30^{+0.08}_{-0.08}$ \\ & ~ & $kT_2$ & ~ & --- & ~~~ &
$0.84^{+0.04}_{-0.03}$ & ~~~ & $0.92^{+0.07}_{-0.06}$ & ~~~ &
$0.79^{+0.05}_{-0.05}$ \\ & ~ & $\Delta N_{\rm H}$ & ~ & --- & ~~~ &
$5.0^{+0.6}_{-0.8}$ & ~~~ & $3.78^{+0.83}_{-1.09}$ & ~~~ &
$4.88^{+0.72}_{-0.86}$ \\ & ~ & $\Gamma$ & ~ & --- & ~~~ & --- & ~~~ &
--- & ~~~ & $1.40^{+0.37}_{-0.46}$ \\ & ~ & $A_1$ & ~ & --- & ~~~ &
--- & ~~~ & --- & ~~~ & $137^{+159}_{-55}$ \\ & ~ & $\chi^2$/DOF & ~ &
2507/1152 & ~~~ & 1950/1149 & ~~~ & 1558/1145 & ~~~ & 1468/1143 \\
\hline &&&&&&&&&& \\ & ~ & $kT$ & ~ & $1.32^{+0.02}_{-0.03}$ & ~~~ &
$1.77^{+0.07}_{-0.07}$ & ~~~ & --- & ~~~ & $1.54^{+0.07}_{-0.06}$ \\ &
~ & $Z$ & ~ & $0.40^{+0.04}_{-0.03}$ & ~~~ & $1.25^{+0.22}_{-0.17}$ &
~~~ & --- & ~~~ & $0.98^{+0.22}_{-0.15}$ \\ NGC 1399 & ~ & $N_{\rm H}$
& ~ & $0.55^{+0.11}_{-0.11}$ & ~~~ & $0.38^{+0.25}_{-0.23}$ & ~~~ &
--- & ~~~ & $0.44^{+0.28}_{-0.26}$ \\ & ~ & $kT_2$ & ~ & --- & ~~~ &
$0.78^{+0.04}_{-0.03}$ & ~~~ & --- & ~~~ & $0.71^{+0.04}_{-0.03}$ \\ &
~ & $\Delta N_{\rm H}$ & ~ & --- & ~~~ & $2.69^{+0.79}_{-0.91}$ & ~~~
& --- & ~~~ & $3.57^{+0.90}_{-1.02}$ \\ & ~ & $\Gamma$ & ~ & --- & ~~~
& --- & ~~~ & --- & ~~~ & $0.86^{+0.57}_{-0.63}$ \\ & ~ & $A_1$ & ~ &
--- & ~~~ & --- & ~~~ & --- & ~~~ & $5.87^{+9.73}_{-4.05}$ \\ & ~ &
$\chi^2$/DOF & ~ & 1280/792 & ~~~ & 941.8/789 & ~~~ & --- & ~~~ &
878.6/787 \\ \hline &&&&&&&&&& \\ & ~ & $kT$ & ~ &
$2.70^{+0.02}_{-0.03}$ & ~~~ & $3.32^{+0.10}_{-0.10}$ & ~~~ &
$3.87^{+0.13}_{-0.21}$ & ~~~ & $3.27^{+0.26}_{-0.12}$ \\ & ~ & $Z$ & ~
& $1.20^{+0.05}_{-0.04}$ & ~~~ & $0.97^{+0.05}_{-0.05}$ & ~~~ &
$0.84^{+0.02}_{-0.06}$ & ~~~ & $0.86^{+0.03}_{-0.07}$ \\ NGC 4696 & ~
& $N_{\rm H}$ & ~ & $1.09^{+0.05}_{-0.04}$ & ~~~ &
$1.44^{+0.06}_{-0.11}$ & ~~~ & $0.89^{+0.22}_{-0.15}$ & ~~~ &
$0.96^{+0.22}_{-0.26}$ \\ & ~ & $kT_2$ & ~ & --- & ~~~ &
$1.40^{+0.02}_{-0.04}$ & ~~~ & $1.52^{+0.04}_{-0.05}$ & ~~~ &
$1.46^{+0.07}_{-0.06}$ \\ & ~ & $\Delta N_{\rm H}$ & ~ & --- & ~~~ &
$0.00^{+0.36}_{-0.00}$ & ~~~ & $0.71^{+0.53}_{-0.60}$ & ~~~ &
$0.59^{+1.01}_{-0.54}$ \\ & ~ & $\Gamma$ & ~ & --- & ~~~ & --- & ~~~ &
--- & ~~~ & $0.76^{+0.59}_{-0.65}$ \\ & ~ & $A_1$ & ~ & --- & ~~~ &
--- & ~~~ & --- & ~~~ & $22.0^{+62.1}_{-16.4}$ \\ & ~ & $\chi^2$/DOF &
~ & 2124/755 & ~~~ & 1219/752 & ~~~ & 990.7/748 & ~~~ & 954.5/746 \\
&&&&&&&&&& \\ \hline &&&&&&&&&& \\
\end{tabular}
\end{center}
 
\parbox {7in} {The best-fit parameter values and 90 per cent ($\Delta
\chi^2 = 2.71$) confidence limits from the analysis of the ASCA
spectra for the central cluster galaxies. Temperatures ($kT$),
metallicities ($Z$), column densities ($N_{\rm H}$), intrinsic column
densities ($\Delta N_{\rm H}$), photon indices ($\Gamma$), power-law
normalizations ($A_1$), and the normalizations of the cooler plasma
emission components in spectral models D, F, G were linked to take the
same values in all detectors. Only the normalizations of the hotter
plasma emission components were allowed to vary independently for each
detector. Temperatures are quoted in keV and metallicities as a
fraction of the solar photospheric value (Anders \& Grevesse
1989). Column densities and intrinsic column densities are in units of
$10^{21}$ atom cm$^{-2}$.  Power-law normalizations are in units of
$10^{-5}$ photon keV$^{-1}$cm$^{-2}$s$^{-1}$ at 1 keV.}
\end{table*}

\addtocounter{table}{-1}
\begin{table*}
\begin{center}
\caption{Spectral results for the Virgo ellipticals}
\vskip 0.1truein
\begin{tabular}{ c c c c c c c c c c c }
\hline & ~ & Parameters & ~ & Model B & ~~~ & Model D &
~~~ & Model F & ~~~ & Model G \\ \hline &&&&&&&&&& \\ & ~ & $kT$ & ~ &
$1.03^{+0.02}_{-0.03}$ & ~~~ & $1.66^{+0.12}_{-0.10}$ & ~~~ & --- &
~~~ & $1.31^{+0.09}_{-0.06}$ \\ & ~ & $Z$ & ~ & $0.26^{+0.02}_{-0.02}$
& ~~~ & $1.34^{+0.33}_{-0.24}$ & ~~~ & --- & ~~~ &
$1.11^{+0.36}_{-0.23}$ \\ NGC 4472 & ~ & $N_{\rm H}$ & ~ &
$0.56^{+0.15}_{-0.14}$ & ~~~ & $0.61^{+0.56}_{-0.52}$ & ~~~ & --- &
~~~ & $1.65^{+0.34}_{-0.69}$ \\ & ~ & $kT_2$ & ~ & --- & ~~~ &
$0.77^{+0.02}_{-0.03}$ & ~~~ & --- & ~~~ & $0.71^{+0.03}_{-0.03}$ \\ &
~ & $\Delta N_{\rm H}$ & ~ & --- & ~~~ & $1.10^{+1.11}_{-0.95}$ & ~~~
& --- & ~~~ & $0.09^{+1.12}_{-0.09}$ \\ & ~ & $\Gamma$ & ~ & --- & ~~~
& --- & ~~~ & --- & ~~~ & $0.78^{+0.55}_{-0.65}$ \\ & ~ & $A_1$ & ~ &
--- & ~~~ & --- & ~~~ & --- & ~~~ & $3.82^{+5.88}_{-2.57}$ \\ & ~ &
$\chi^2$/DOF & ~ & 1006/375 & ~~~ & 536.7/372 & ~~~ & --- & ~~~ &
474.0/370 \\ \hline &&&&&&&&&& \\ & ~ & $kT$ & ~ &
$0.661^{+0.005}_{-0.006}$ & ~~~ & $0.723^{+0.030}_{-0.021}$ & ~~~ &
$0.703^{+0.014}_{-0.014}$ & ~~~ & $0.677^{+0.010}_{-0.009}$ \\ & ~ &
$Z$ & ~ & $0.32^{+0.02}_{-0.02}$ & ~~~ & $0.35^{+0.03}_{-0.02}$ & ~~~
& $0.43^{+0.03}_{-0.04}$ & ~~~ & $0.62^{+0.13}_{-0.09}$ \\ NGC 4636 &
~ & $N_{\rm H}$ & ~ & $1.42^{+0.10}_{-0.11}$ & ~~~ &
$1.50^{+0.12}_{-0.11}$ & ~~~ & $1.24^{+0.11}_{-0.13}$ & ~~~ &
$1.03^{+0.26}_{-0.25}$ \\ & ~ & $kT_2$ & ~ & --- & ~~~ &
$0.550^{+0.041}_{-0.053}$ & ~~~ & $0.509^{+0.038}_{-0.046}$ & ~~~ &
$0.387^{+0.052}_{-0.057}$ \\ & ~ & $\Delta N_{\rm H}$ & ~ & --- & ~~~
& $0.00^{+0.03}_{-0.00}$ & ~~~ & $0.00^{+0.23}_{-0.00}$ & ~~~ &
$2.45^{+2.00}_{-1.47}$ \\ & ~ & $\Gamma$ & ~ & --- & ~~~ & --- & ~~~ &
--- & ~~~ & $1.49^{+0.25}_{-0.25}$ \\ & ~ & $A_1$ & ~ & --- & ~~~ &
--- & ~~~ & --- & ~~~ & $10.4^{+4.7}_{-3.3}$ \\ & ~ & $\chi^2$/DOF &
~ & 1512/235 & ~~~ & 1444/232 & ~~~ & 1060/229 & ~~~ & 475.0/227 \\
\hline &&&&&&&&&& \\ & ~ & $kT$ & ~ & $0.96^{+0.04}_{-0.04}$ & ~~~ &
$2.51^{+0.23}_{-0.20}$ & ~~~ & --- & ~~~ & $1.75^{+0.22}_{-0.26}$ \\ &
~ & $Z$ & ~ & $0.19^{+0.02}_{-0.03}$ & ~~~ & $1.01^{+0.38}_{-0.28}$ &
~~~ & --- & ~~~ & $0.66^{+0.32}_{-0.21}$ \\ NGC 4649 & ~ & $N_{\rm H}$
& ~ & $0.81^{+0.30}_{-0.27}$ & ~~~ & $0.00^{+0.13}_{-0.00}$ & ~~~ &
--- & ~~~ & $0.00^{+0.26}_{-0.00}$ \\ & ~ & $kT_2$ & ~ & --- & ~~~ &
$0.79^{+0.04}_{-0.03}$ & ~~~ & --- & ~~~ & $0.77^{+0.05}_{-0.04}$ \\ &
~ & $\Delta N_{\rm H}$ & ~ & --- & ~~~ & $2.11^{+0.62}_{-0.70}$ & ~~~
& --- & ~~~ & $2.34^{+0.66}_{-0.80}$ \\ & ~ & $\Gamma$ & ~ & --- & ~~~
& --- & ~~~ & --- & ~~~ & $0.55^{+0.72}_{-0.54}$ \\ & ~ & $A_1$ & ~ &
--- & ~~~ & --- & ~~~ & --- & ~~~ & $2.24^{+5.52}_{-1.45}$ \\ & ~ &
$\chi^2$/DOF & ~ & 606.3/240 & ~~~ & 318.8/237 & ~~~ & --- & ~~~ &
291.1/235 \\ &&&&&&&&&& \\ \hline
\end{tabular}
\end{center}
\parbox {7in}
{}
\end{table*}

The results from the spectral modeling are summarized in Table 3. We
see that the two-temperature model, model D, invariably provides a
much better description of the ASCA spectra for the galaxies than the
single-temperature model (model B), with a typical reduction in
$\chi^2$ of a few hundred for the introduction of only three
additional degrees of freedom in the fit. For NGC 1399, 4472 and 4649,
the use of the two temperature model results in a significant increase
in the inferred metallicity of the X-ray gas in the galaxies, from
values of $0.2-0.4$ solar with spectral model B, to values consistent
with (or slightly exceeding) the solar value, with spectral model
D. This is in agreement with the results previously-reported by Buote
\& Fabian (1998; see also Buote 1999).

For M87, NGC 4696 and NGC 4636, we found that the fits were further
significantly improved by allowing the abundances of various
individual elements to be included as additional free parameters in
the fits.  (With spectral model D, all elements are linked to vary in
the same ratio, relative to their solar values). Improvements in the
measured $\chi^2$ values of several hundred were obtained for these
objects by allowing the abundances of Mg, Si and S to be included as
free parameters in the fits (with only a single extra degree of
freedom being associated with each extra element included as a free
fit parameter). The results on the individual element abundances for
M87 and NGC 4696 are discussed by Allen \etal (1999). The results for
NGC 4636 are detailed in Section 6.

\subsection{The requirement for the power-law components}

\begin{figure}
\hspace{-0.3cm}\psfig{figure=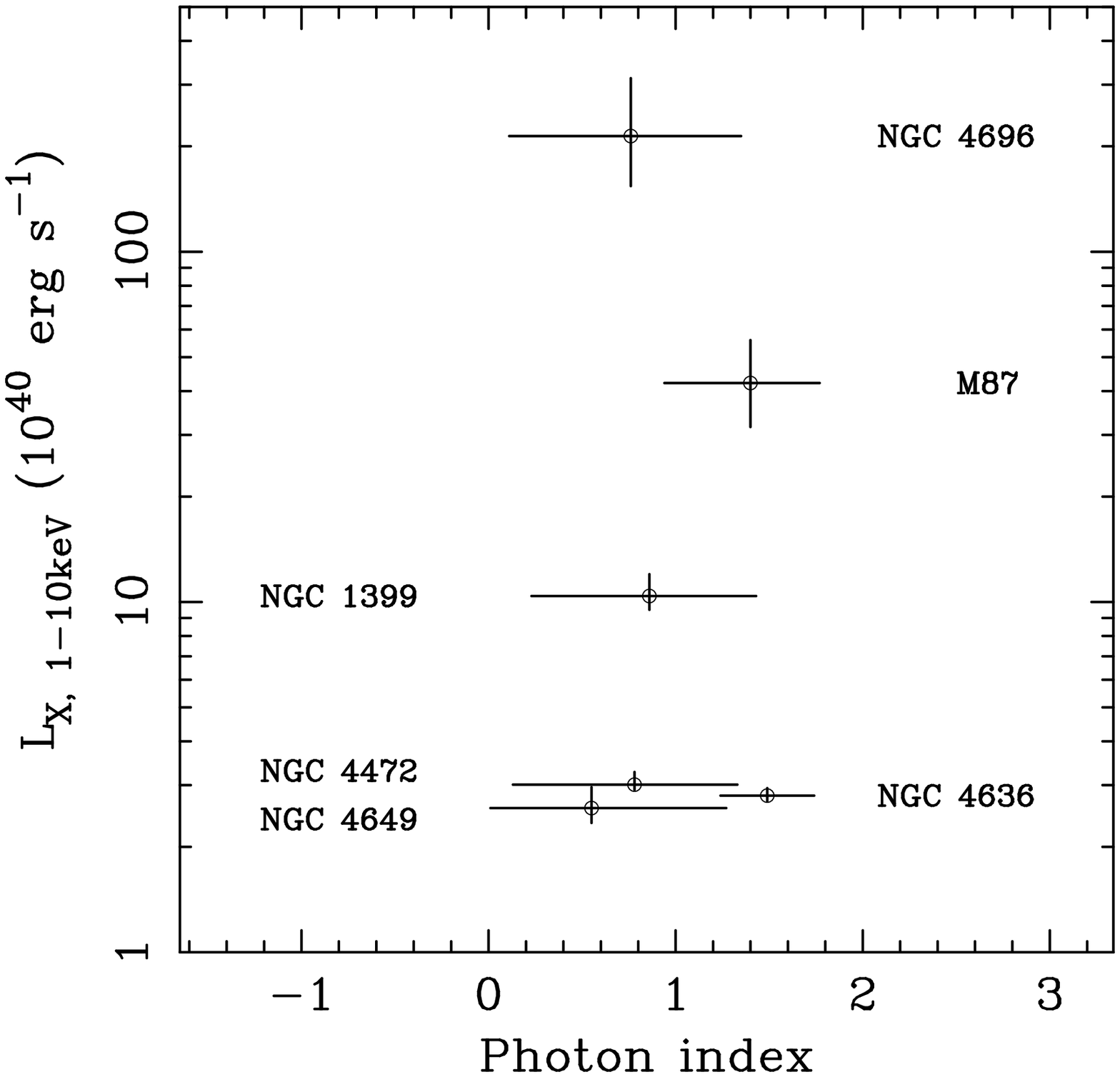,width=0.65\textwidth,angle=270}
\caption{The intrinsic $1-10$ keV luminosities of the detected
power-law components as a function of the observed photon index,
$\Gamma$. Error bars are 90 per cent ($\Delta \chi^2 = 2.71$)
confidence limits. No correlation between the luminosity and photon
index is observed.}
\end{figure}

The results on the power-law components detected in the ASCA spectra
are summarized in Table 4. In all cases the introduction of the
power-law component into the two-temperature models (models D and F)
leads to a highly significant improvement in the fit. For guidance, a
reduction in $\chi^2$ of $\Delta \chi^2 \sim 10$ with the introduction
of the power-law component (2 extra fit parameters) is significant at
approximately the 99 per cent confidence level (for a fit with 1000
degrees of freedom and a reduced $\chi^2$/DOF value $\sim 1.0$). The
observed improvements range from $\Delta \chi^2 \sim 30$ to $\Delta
\chi^2 \sim 600$.

In all cases the slopes of the power-law components are significantly
flatter ($\Gamma = 0.6-1.5$) than the canonical value of $\Gamma \sim
1.8$ obtained for Seyfert galaxies (\eg Nandra \etal 1997, Reynolds
1997). The weighted mean best-fit photon index for the sources in our
sample is $\Gamma = 1.22$. The observed $2-10$ keV fluxes range from
$5.7 \times 10^{-13}$ to $8.7 \times 10^{-12}$ \ergpcmsqps, and the
intrinsic $1-10$ keV luminosities of the power-law components
(corrected for Galactic and intrinsic absorption as determined with
spectral model G, and calculated using the luminosity distances listed
in Table 1) range from $2.6 \times 10^{40}$ (NGC 4649) to $2.2 \times
10^{42}$ \ergps (NGC 4696). These luminosities are also less than
those typically associated with Seyfert nuclei. Figure 1 shows the
intrinsic $1-10$ keV luminosities as a function of the observed photon
index. No clear correlation between the two parameters is
observed. Fig. 2 shows the ASCA spectrum and best-fitting model for
M87 (spectral model G). For clarity, only the results for the S0
detector are shown.

The detection of hard, power-law emission components from all six
galaxies is not easily explained as an artifact of the
analysis. Applying the same method to an ASCA observation of NGC 1275,
the dominant galaxy of the nearby ($z=0.0183$) Perseus Cluster, which
is known to contain an active nucleus, we determine a photon index for
the nuclear emission of $\Gamma = 2.05\pm0.05$, in good agreement with
the typical values determined for such galaxies (Nandra \etal 1997;
Reynolds 1997; Turner \etal 1997).  The $2-10$ keV flux associated
with the nucleus of NGC 1275, $F_{\rm X,2-10} \sim 1.8 \times
10^{-10}$ \ergpcmsqps (implying an intrinsic $1-10$ keV luminosity of
$\sim 4 \times 10^{44}$ \ergps), is also significantly larger than the
fluxes associated with the harder power-law components detected in the
present sample of objects. Secondly, when we apply the same analysis
method (using spectral model D) to an observation of the central
regions of the Coma Cluster ($z=0.0232$), where the observations are
centred on the X-ray centroid of the cluster rather than an individual
galaxy (the cluster does not contain a single, optically-dominant
galaxy in its core), we find no improvement in the fit ($\Delta \chi^2
=0.0$) on the introduction of a hard, power-law component. The upper
(90 per cent confidence) limit to the $2-10$ keV flux of any power-law
component with a photon index $\Gamma = 2.0$ in the ASCA data for
centre of the Coma Cluster using spectral model D is $F_{\rm X,2-10} < 2.2 
\times 10^{-12}$ \ergpcmsqps, significantly less than the measured values for M87 and
NGC 4696. We also note that the application of a similar analysis
procedure to ASCA observations of more distant, luminous cooling-flow
clusters does not, generally, indicate the presence of hard ($\Gamma
\sim 1.2$) power-law components in these systems. (We do not expect to
detect hard, power-law components with luminosities comparable to
those reported here in clusters an order of magnitude (or more) more
X-ray luminous than the Virgo or Centaurus clusters; {\it c.f.}
Section 5.3) This further suggests that the hard, power-law components
detected in the present sample of elliptical galaxies are not
artifacts due to systematic errors in the modeling of diffuse
cluster/galaxy emission in these systems. (A more detailed discussion
of the ASCA data for the Perseus and Coma clusters is presented by
Allen \etal 1999.  For a discussion of more distant, luminous
cooling-flow clusters see Allen 1999).

Finally, we note that the data for NGC 4636 exhibit a systematic
discrepancy at low energies ($E \approxlt 1$keV) between the S0 and S1
detectors, which contributes significantly to the measured $\chi^2$. (This
discrepancy is also noted by Buote 1999). Ignoring the data in this region 
and repeating the analysis with spectral model G leads to a lower 
$\chi^2$ value ($\chi^2 = 284$ for 201 degrees of freedom) and constraints on the 
power-law component in good agreement with those listed in Table 2.

\subsection{A bremsstrahlung model for the hard X-ray emission}

We have investigated whether the hard X-ray emission from the elliptical
galaxies can also be parameterized by a simple bremsstrahlung model. 
To do this, the power-law component in spectral model G was replaced with 
a thermal bremsstrahlung component, with the temperature, $kT_{\rm brem}$, and 
normalization, $A_2$, as fit parameters. The results from the fits with
this modified model are summarized in Table 5. The $\chi^2$ values obtained
indicate that the bremsstrahlung model provides as good a
parameterization for the hard X-ray emission from the galaxies
as the power-law model. Generally, the results constrain the 
temperatures of the bremsstrahlung components to be $\geq 10$ keV, 
with the upper limits on the temperatures unconstrained. (ASCA 
data in the $0.6-10.0$ keV band cannot reliably constrain the 
temperatures of plasmas hotter than $\sim 10$ keV). The results on the other 
fit parameters are essentially identical to those listed in Table 3. The 
fluxes and intrinsic luminosities associated with the bremsstrahlung 
components are also in good agreement with the values listed in Table 4,
using the power-law model. The measured temperatures for the bremsstrahlung 
components are significantly hotter than those of typical Galactic binary 
sources (\eg Christian \& Swank 1998) and support the interpretation for 
the origin of the hard X-ray emission from the galaxies in terms of low 
radiative efficiency accretion flows, given in Section 4.

We have examined whether the statistical quality of the fits 
with spectral model G (or the modified model G, incorporating 
the bremsstrahlung component) can be improved by the addition of 
further model components. In all cases, the introduction of a further
power-law or bremsstrahlung component did not significantly improve the fits. 
The introduction of a third, thermal plasma component (as described in
Section 3.1) also leads to no significant improvement (at the 99 per cent 
confidence level) for any object, except NGC 4696. For this source, the 
introduction of a third plasma component (with the element abundances linked 
to be equal to those in the other two components) leads to a drop in 
$\chi^2$ of $\Delta \chi^2 = 37$, and slightly modified constraints on the 
power-law emission ($\Gamma = 1.10^{+0.36}_{-0.44}$ and $A_1 = 
5.8^{+6.2}_{-3.5} \times 10^{-4}$ photon keV$^{-1}$cm$^{-2}$s$^{-1}$). We
note that the relatively high temperature ($kT \sim 3.3$ keV) 
for the hotter thermal component in NGC 4696 simply reflects the virial 
temperature of the host cluster. NGC 4696 is the most distant galaxy
included in the present study and the 3.7 arcmin (66 kpc) S1 aperture
used includes significant flux from the extended cluster gas. 

Finally, we note that several of the $\chi^2$ values listed in Tables 
$3-5$ (in particular, those for the galaxies with the highest 
signal-to-noise ratios in their ASCA spectra) indicate that the best-fit 
models are, formally, statistically unacceptable. However, the high 
$\chi^2$ values obtained are primarily due to residual systematic errors in 
the instrument response matrices and plasma emission models (although we do
not expect these to significantly effect the main conclusions reported here.)

\section{The origin of the hard X-ray emission}

\subsection{Accretion processes}


The presence of hard, power-law X-ray emission from astrophysical
sources provides a discriminating signature of accretion processes
around black holes. Such emission is usually attributed to the
presence of a hot, tenuous coronal plasma (probably magnetically)
coupled to an accretion disk or a hot (ADAF-type) accretion flow, but
can also arise in the shock sites of a jet. Other astrophysical
processes tend to give rise to softer X-ray emission, with a steeper
photon index (when this emission is modeled as a simple
power-law). The flat slopes of the power-law components detected in
our sample of elliptical galaxies, together with the dynamical
evidence for supermassive black holes with masses of
$10^8-10^{10}$\Msun~in the nuclei of these objects, argues strongly
for this emission being intimately related to the accretion process.

The relatively broad point spread function of the ASCA mirrors
(Section 2.2) does not allow us to resolve the X-ray emission from the
galaxy cores into more than a single integrated spectrum. We therefore
cannot separate the X-ray emission associated with the jets in these
objects (\eg knot A in M87, which previous ROSAT High Resolution
Imager observations have resolved; Reynolds \etal 1996, Harris, Biretta \& 
Junor 1997, 1998) from the accretion disks themselves. However, the fact 
that the galaxy with
the strongest radio emission, M87, in which the radio and X-ray jet
emission is relativistically beamed towards us, exhibits a
significantly weaker hard X-ray component than NGC 4696
(which has a twin lobe radio structure and a total 4.85 GHz radio flux
$\sim 40$ times lower than M87) at least suggests that the jet
emission does not dominate the detected hard X-ray flux. This is
further supported by the detections of hard, power-law X-ray
components, with similar characteristic slopes, in objects such as NGC
4636, 4649, and 4472, in which the radio activity is at the level of
a mJy or less. Sharp cut-offs observed in the radio spectra of these
objects (Di Matteo \etal 1999a) also strongly constrain the populations
of non-thermal particles responsible for synchrotron radiation and
synchrotron self-Compton emission (in X-rays) from a jet and/or
outflow associated with a low radiative-efficiency accretion flow
(although contributions from a non-thermal distribution of
relativistic electrons might still be important in the central cluster
galaxies with more dominant radio activity; Di Matteo \etal 1999b).

\subsubsection{The effects of intrinsic absorption}

The ASCA spectra alone cannot reliably discriminate between 
whether the observed power-law components are intrinsically flat or have 
steeper photon indices which have been modified by the effects of intrinsic 
absorption over and above that accounted for by spectral model G. 
(This uncertainty is primarily due to the complex spectrum of the diffuse, 
thermal emission at soft X-ray energies). In general, the inclusion of an 
extra $\sim 5 \times 10^{22} - 10^{23}$ \apc~of intrinsic absorption 
associated with the power-law components leads to power-law 
photon indices of $\Gamma \sim 2$, and provides similarly good fits to 
the ASCA data (although the determination of confidence limits on the 
fit parameters becomes difficult with such models). Conversely, if the 
true intrinsic column densities acting on the hard X-ray components are 
overestimated using spectral model G, then the intrinsic photon indices of 
these components will actually be flatter than the values listed in Table 4.

An intrinsic photon index of $\Gamma \sim 1.8-2$ is typically observed
in AGN (\eg Nandra \etal 1997; Reynolds 1997; Turner \etal 1997),
where the accretion is thought to occur via a thin disk, and where the
radiative efficiency is $\sim 10$ per cent. However, as discussed in
Section 1, if the accretion in the present sample of elliptical
galaxies were to proceed in this manner, the X-ray luminosities
associated with their power-law emission components should be $3-5$ orders of
magnitude larger than the observed values.  There is no simple way in
which to modify the observed X-ray fluxes (particularly when one takes
into account the joint ASCA and ROSAT results; Section 5.1) and
measured photon indices to agree with the predictions from the
standard thin-disk accretion models by simply including extra
absorption on the power-law components. 

IRAS measurements of the 60 and 100$\mu m$ fluxes from the galaxies
(determined using the IPAC SCANPI software applied to co-added IRAS
scans) also constrain the X-ray luminosities that may be
absorbed and reprocessed to infrared wavelengths in these systems.
For M87, the observed 60 and 100$\mu m$ luminosities ($L_{\rm 60\mu
m}\sim 8\pm1 \times 10^{41}$\ergps~and $L_{\rm 100\mu m}\sim 4\pm1
\times 10^{41}$\ergps, respectively) are comparable to the $1-10$ keV
luminosity of the hard X-ray component (Table 4). The infrared 
luminosities for the other galaxies are summarized in
Table 6. In general, the observed 100$\mu m$ luminosities do not exceed the
$1-10$ keV values by more than an order of magnitude (with the 
exception of NGC 4649, for which the observed 100$\mu m$ luminosity is 
$\sim 500$ times larger than the $1-10$keV value) and the presence of an 
absorbed active nucleus with an X-ray luminosity $4-5$ orders of magnitude more
luminous in X-rays than the values listed in Table 4 can be firmly ruled
out. We note, however, that although the X-ray and infrared data argue
strongly against the observed hard, power-law components being due to
intrinsically absorbed, steep ($\Gamma \sim 2$) X-ray spectra with 
luminosities $\sim 10^{46}$\ergps, the very flat photon indices 
observed in some galaxies (with $\Gamma < 1$) suggest that intrinsic 
absorption may play some role. Such flat slopes could result from 
absorption of few $\times 10^{21}$\apc~ acting on emission spectra 
with intrinsic photon indices, $\Gamma \sim 1.4$ (which could be 
produced by low-radiative efficiency accretion flows with electron 
temperatures $\approxlt 100$keV; Section 4.1.2). This is consistent with the 
observation that M87, which has an X-ray photon index, $\Gamma =1.4$, 
exhibits a lower $L_{\rm 100\mu m}/L_{\rm X,1-10}$ ratio
than those galaxies with flatter X-ray slopes.  

\begin{figure*}
\hbox{
\hspace{0cm}\psfig{figure=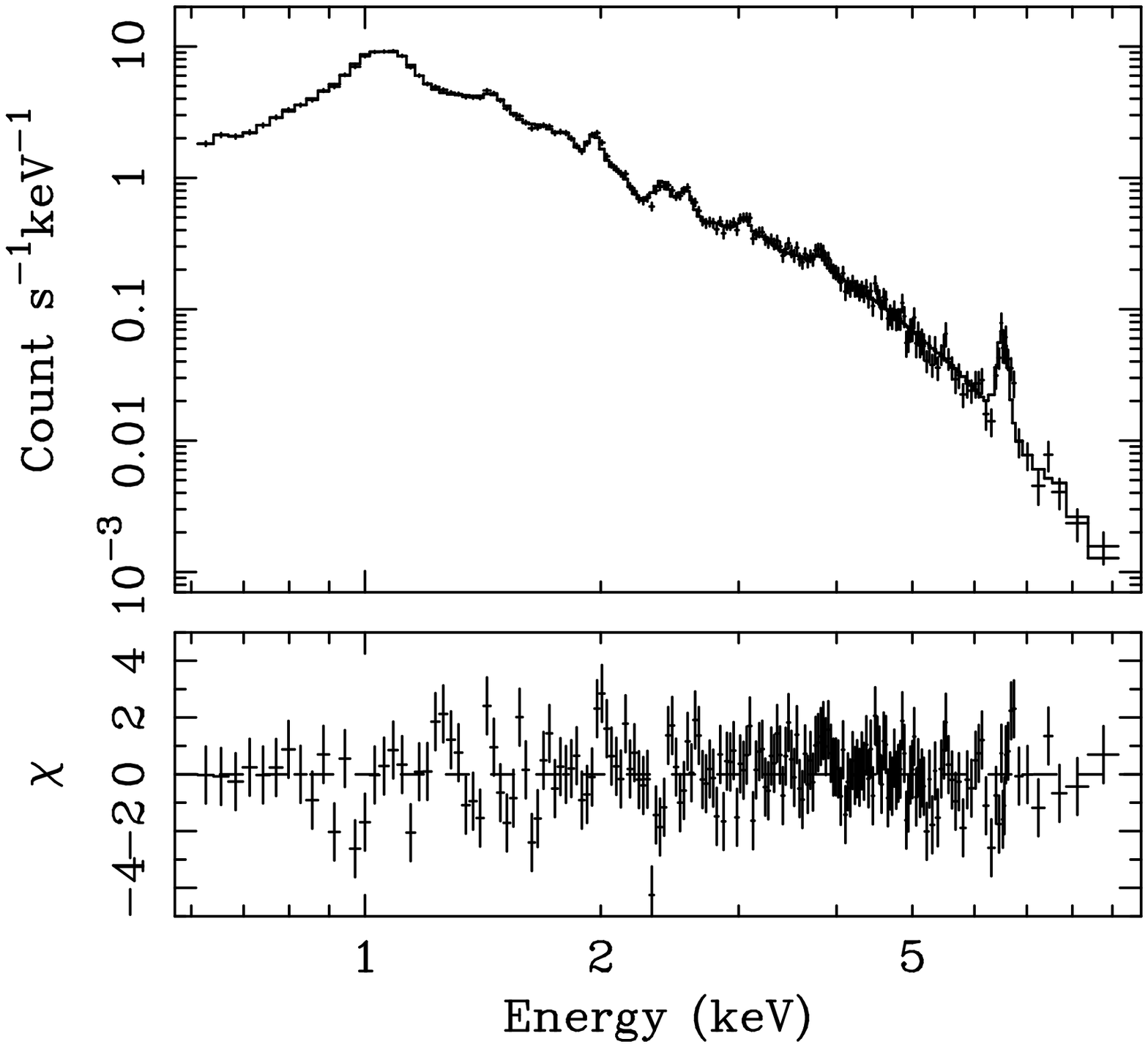,width=0.65\textwidth,angle=270}
\hspace{-2.5cm}\psfig{figure=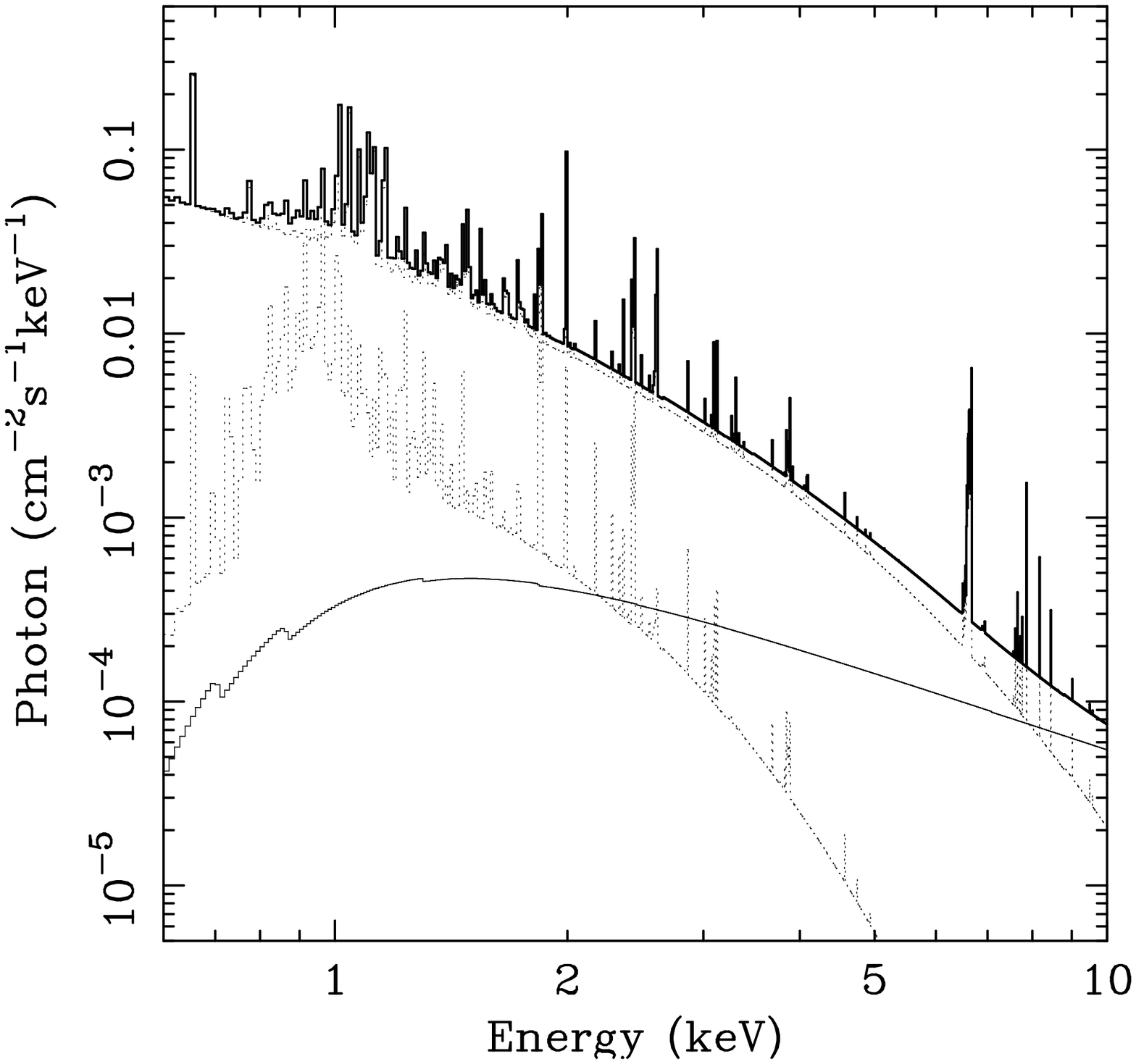,width=0.567\textwidth,rheight=7.91cm,rwidth=5cm,angle=270}
}
\caption{(Left) The ASCA S0 data and best-fitting
two-temperature-plus-power-law model for M87 (with the abundances of
various individual elements also included as free fit parameters;
Allen \etal 1999) and the residuals to the fit in units of
$\chi$. (Right) The best-fit model for the S0 data, showing the
relative contributions of the hot and cool plasma emission (dotted
curves) and power-law components (lower solid curve) to the total flux
(bold solid curve).}
\end{figure*}

\subsubsection{Low radiative efficiency accretion flows}

Although the observed properties of the elliptical nuclei are not
easily explained within the context of unified models for Seyfert
galaxies, the characteristic hard X-ray, radio and (lack of) optical
emission from these sources (i.e. the typically non-thermal character
of their spectra) can be accounted for by models of low-radiative
efficiency accretion (such as an ADAF; Narayan \& Yi 1994, 1995)
coupled with outflows/winds (Blandford \& Begelman 1999), with the
accretion occurring at approximately the Bondi rates.

In an advection dominated flow, which occurs at accretion rates $\Mdot
\approxlt \alpha^2 \Mdot_{\rm Edd}$ (where $\Mdot_{\rm Edd}$ is the
accretion rate at the Eddington limit) the viscosity (parameterized by
the constant $\alpha \approxgt 0.1$) is assumed to dissipate most of the energy
locally into ions. The ions cannot cool (Coulomb scattering between
the ions and electrons is very inefficient in the low density plasma;
no other form of electron-ion coupling is assumed and the gas is
two-temperature) and flow inwards, carrying an increasing amount of
thermal energy. In the absence of an energy sink, which would normally
be present in the form of efficient radiation in a thin disk, the gas
will be unbound (i.e. have a Bernoulli parameter greater than zero)
unless the binding energy is transported radially outward by the
viscous torque, in the form of a wind (Blandford \& Begelman
1998). Within the context of low-radiative efficiency accretion flows
(unavoidably coupled with outflows), most of the accreted mass,
angular momentum and energy will be lost at large distances in the
flows and the central pressures and densities will be much reduced.

In these more generalized models for low-radiative efficiency
accretion flows, the density varies as $\rho \propto r^{-3/2+p}$
(where the accretion rate $\Mdot \propto r^{p}$, with $p\sim 1$) such
that the emission in the central regions is highly suppressed. The
thermal electrons, at a temperature of $\sim 100 \keV$, radiate by
synchrotron emission, inverse Compton scattering (of synchrotron photons) and
bremsstrahlung emission.  Most of the synchrotron emission and its
Comptonization will occur in the innermost regions (within a few
Schwarschild radii) of the flow (where the temperature is the high
enough, $\approxgt 10^9 \K$, for these processes to become important)
and will therefore be highly suppressed in the presence of winds
\footnote{Although the luminosities due to both bremsstrahlung and
synchrotron processes vary as the square of the accretion rate
({\it i.e.} the density), in the presence of a wind the accretion rate will
be much reduced in the inner regions where most of the synchrotron
emission is produced. More importantly, in the very low density inner
regions the electron temperature profile is much flatter; the
synchrotron power $\propto T^7$ is highly suppressed (Di Matteo \etal
1999b; Quataert \& Narayan 1999). Comptonization of this
component is also highly suppressed, its importance also depending on
the scattering optical depth $\tau \propto \Mdot$, which always
remains low in the presence of a wind.}  (as emphasized by the
observations and modeling discussed by Di Matteo \etal 1999a,b; see also
Quataert \& Narayan 1999). Bremsstrahlung emission, in contrast,
arises from all radii in the flow and should, therefore, dominate the
emission from sources in which low radiative efficiency accretion is
associated with winds. The X-ray spectrum due to thermal
bremsstrahlung in the optically thin gas has a very flat spectrum up
to its cut off frequency at about $h\nu \sim kT$. Most of the
bremsstrahlung luminosity, $\propto \rho^2 T^{-1/2} r^3
\exp^{-h\nu/kT}$ with $\rho \propto \alpha^{-1} M_{\rm BH}^{-1}\Mdot
r^{-3/2 +p}$, originates at larger radii where the density is the
highest. The temperature profile in an ADAF with winds is virtually
constant with radius with $kT \approx100-200 \keV$ for $r\approxlt$ a few
hundred Schwarschild radii, and tracks the virial temperature in the outer
regions.  Even in the presence of a very strong wind the bremsstrahlung
spectrum would not be highly suppressed up to energies $kT
\approxlt 10\keV$, as the minimum temperature in the outer region of
the flow is $\sim 10^{12}/r_{\rm out}$K and the outer radius $r_{\rm
out}$ is of the order of a $10^{3}-10^{4}$ Schwarschild radii.  As
discussed by Di Matteo \etal (1999b), the rates at which material is
required to be fed from the hot interstellar medium to the outer regions of 
the accretion flows, to explain the luminosities of the observed hard,
X-ray components, are consistent with the expected Bondi accretion estimates of 
$0.1-1 \Msun$yr$^{-1}$.  The emission from such a flow would
be dominated by a thermal bremsstrahlung at temperatures $\approxlt 100
\keV$ (as expected in the outer regions of the flows) resembling a
hard ($\Gamma \sim 1.4$) power law in the ($1-10$ keV) ASCA band (as required by the
data). The differences in luminosity between the objects in our sample
can then be ascribed to differences in the black hole masses and Bondi
accretion rates, with the hard, power-law components in the central
cluster galaxies being more luminous due to their higher density
environments. This is illustrated in Fig. 3 where we show the
intrinsic ($1-10$ keV) luminosities of the hard, power-law components
as a function of the bolometric luminosity of the X-ray gas within a
radius of 10 kpc in the galaxies. A clear correlation between these
quantities is observed, with NGC 4696, the dominant galaxy of the
Centaurus Cluster and the galaxy with the most luminous hard,
X-ray component, also exhibiting the largest X-ray gas luminosity
within a radius of 10 kpc. Detailed modeling and discussion of these
issues, and of the effects of non-thermal particle distributions in
the winds and/or jets, are presented by Di Matteo \etal (1999b).

\subsection{Other sources of X-ray emission}

Although the observed properties of the hard X-ray components can be
accounted for by models of low-radiative efficiency accretion onto the
central supermassive black holes in the galaxies, a variety of other
sources (given the large field of view of the ASCA instruments) may
contribute to the integrated X-ray spectra.

\subsubsection{Binary X-ray sources}

Undoubtedly, some contribution towards the harder X-ray emission from
the galaxies will be made by binary X-ray sources (\eg Canizares,
Fabbiano \& Trinchieri 1987; Fabbiano 1989).  Bright Galactic
X-ray binaries and black-hole candidates exhibit persistent X-ray
luminosities in the range $10^{36}-10^{38}$ \ergps~(\eg White \etal
1988). The luminosities associated with the power-law components detected 
in the Virgo ellipticals could
therefore be accounted for by a few $10^2-10^4$ such sources (or $\sim
10^4-10^6$ sources for the more luminous central cluster galaxies). The 
luminosities of the power-law components in NGC 1399, 4472, 4636 and 4649 
are consistent with an extrapolation of the $L_{\rm X}/L_{B}$ relation 
determined for irregular and spiral galaxies (Fabbiano \& Trinchieri 1985), 
in which the X-ray emission is dominated by discrete sources (binary X-ray 
sources, stars and central active nuclei). Thus, binary X-ray sources
could contribute at a significant level (comparable to the accretion 
flows; see below) to the harder X-ray emission from these galaxies. 
For M87 and NGC 4696, however, the X-ray luminosities of the hard, power-law 
components are an order or magnitude 
larger than the values estimated from the $L_{\rm X}/L_{B}$ relation, based 
on their observed blue luminosities. Binary X-ray sources are therefore
unlikely to contribute significantly to the 
hard, power-law components detected in M87 and NGC 4696. (The result for
M87 is strongly supported by the ROSAT HRI observations discussed in Section
5.1). 

In addition, although individual binary sources can exhibit 
very hard X-ray spectra, a more typical spectrum in the $2-10$ keV band
would have a photon index (using a simple power-law model) in the range 
$\Gamma = 1.5-2.5$ or, using a thermal bremsstrahlung model, a
temperature, $kT \sim 4-7$ keV (\eg Fabbiano \etal 1987; White \etal 1988; 
Makishima \etal 1989; Tanaka \etal 1996; Christian \& Swank 1997), with 
little or no associated intrinsic absorption (Fabbiano \etal 1987). These power-law slopes 
are significantly steeper (or, equivalently, the bremsstrahlung temperatures 
are lower) than the observed values for the hard emission
components detected in the present sample of galaxies. (We recall that the 
observed photon indices show no clear correlation with the luminosities of 
the hard, power-law components; Fig. 1). We have examined the effect of 
including a second power-law component, with a fixed photon index of 
$\Gamma = 2.0$, in the analysis of the ASCA data for NGC 4472 and 4636 with 
spectral model G. (We assume that the intrinsic absorption acting on both 
power-law components and the cooler plasma component is the same since the 
ASCA spectra cannot easily constrain more detailed models.) For both systems, 
we find that the fits are not significantly improved by the introduction of 
the steeper power-law components ($\Delta \chi^2 =0.0$; see also Section
3.4), although the maximum (90 per cent confidence) $2-10$ keV fluxes 
associated with these components ($2.0 \times 10^{-13}$ and $3.3 
\times 10^{-13}$ \ergpcmsqps) are approximately 30 and 60 per cent of the 
values listed in Table 4, respectively. With the steeper components included 
at their maximum allowed levels, the photon indices of the harder power-law 
components in NGC 4472 and 4636 are reduced to $0.7^{+0.6}_{-1.1}$ and 
$0.9^{+0.3}_{-0.6}$, respectively. 

We note that the relatively low source fluxes and high 
background count rates at harder energies, together with the 
broad point spread function of the ASCA mirrors, 
prevent us from placing useful constraints, from the ASCA imaging data, 
on the relative contributions to the hard X-ray fluxes from central point 
sources and extended emission components in the galaxies. (For M87 and 
NGC 4696, the extended cluster emission dominates the X-ray spectra across 
essentially the entire ASCA energy band; Section 5.3). Future observations 
with the Chandra Observatory will provide an answer to this question.

We conclude that binary X-ray sources are unlikely to dominate the hard, 
power-law emission detected in M87 and NGC 4696, although they will 
contribute to the measured fluxes in the galaxies with weaker hard, 
X-ray components.

\subsubsection{Cosmic XRB fluctuations}

It is important to consider the effects of fluctuations in the
Cosmic XRB on the results, particularly given the close match between
the weighted mean slope of the detected power-law components ($\Gamma
\sim 1.2$) and the XRB ($\Gamma=1.4$). Following Barcons, Fabian \&
Carrera (1998), we have estimated the confusion noise, $\sigma_{\rm
c}$ (the flux equivalent to a $1-\sigma$ variation in the XRB
intensity histogram) for the ASCA observations. Assuming $\sigma_{\rm
c}=2.0\times 10^{-12} \Omega^{2/3}$, where $\Omega$ is the beam area
of the ASCA SIS spectra in degree$^2$, and normalizing to the GINGA
observations of Butcher \etal (1997), we obtain the confusion limits
listed in column 6 of Table 4. For the strongest sources (M87 and NGC
4696) the detected fluxes are much larger (a factor $40-60$) than the
confusion limits.  For even the weakest sources (\eg NGC 4636), the
detected fluxes are $4-5$ times larger than the confusion limits. We
also recall that the use of an independent on-chip
background-subtraction method in the analysis of the ASCA observations
made in 4-CCD mode (section 2) provides very similar results on the
power-law components.

\begin{figure}
\hspace{-0.3cm}\psfig{figure=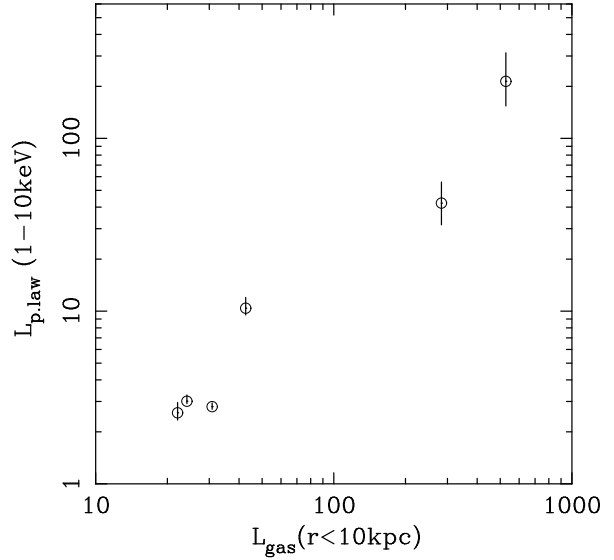,width=0.65\textwidth,angle=270}
\caption{The intrinsic $1-10$ keV luminosities of the hard, power-law
components as a function of the bolometric luminosity of the X-ray gas
within a radius of 10 kpc of the galaxy centres. (All luminosities are in
units of $10^{40}$ \ergps.) A clear correlation between the quantities is 
observed, with a best-fitting slope (determined from a fit with a power-law 
model) of between 1 and 2.}
\end{figure}

\begin{table*}
\begin{center}
\caption{Properties of the hard, power-law components }
\vskip 0.1truein
\begin{tabular}{ c c c c c c c c c c c c }
\hline       
\multicolumn{1}{c}{} &
\multicolumn{1}{c}{} &
\multicolumn{2}{c}{FIT PARAMETERS} &
\multicolumn{1}{c}{} &
\multicolumn{2}{c}{} &
\multicolumn{1}{c}{} &
\multicolumn{1}{c}{CON.} &
\multicolumn{1}{c}{} &
\multicolumn{2}{c}{STATISTICS} \\
&&&&&&&& LIMIT  &&& \\   
        & ~ &   $\Gamma$             &  $A_1$                 & ~ & $L_{\rm X,1-10}$        & $F_{\rm X,2-10}$        & ~~~&  $\sigma_{\rm c,2-10}$   & ~~~ &  $\Delta \chi^2$ & $\chi^2$/DOF \\
\hline                                                                                                                                                             
&&&&&&&&&& \\                                                                                                                                               
M87     & ~ & $1.40^{+0.37}_{-0.46}$ & $137^{+159}_{-55}$     & ~ & $42.2^{+13.7}_{-10.6}$  & $86.9^{+17.0}_{-15.6}$  & ~ & 1.52 & ~ &  90.0 & 1468.1/1143   \\
NGC 1399 & ~ & $0.86^{+0.57}_{-0.63}$ & $5.87^{+9.73}_{-4.05}$ & ~ & $10.4^{+1.6}_{-0.9}$    & $9.26^{+0.30}_{-0.23}$  & ~ & 1.24 & ~ &  63.2 & 878.6/787   \\
NGC 4696 & ~ & $0.76^{+0.59}_{-0.65}$ & $22.0^{+62.1}_{-16.4}$ & ~ & $214^{+99}_{-60}$       & $42.0^{+12.9}_{-9.9}$   & ~ & 1.12 & ~ &  38.8 & 954.7/746     \\
NGC 4472 & ~ & $0.78^{+0.55}_{-0.65}$ & $3.82^{+5.88}_{-2.57}$ & ~ & $3.01^{+0.26}_{-0.10}$  & $7.09^{+0.26}_{-0.08}$  & ~ & 1.05 & ~ &  62.7 & 474.0/370    \\
NGC 4636 & ~ & $1.49^{+0.25}_{-0.25}$ & $10.4^{+4.7}_{-3.3}$   & ~ & $2.80^{+0.14}_{-0.10}$  & $5.72^{+0.16}_{-0.16}$  & ~ & 1.32 & ~ &  585  & 475.0/227   \\
NGC 4649 & ~ & $0.55^{+0.72}_{-0.54}$ & $2.24^{+5.52}_{-1.45}$ & ~ & $2.58^{+0.38}_{-0.24}$  & $6.21^{+0.24}_{-0.36}$  & ~ & 1.08 & ~ &  27.7 & 291.1/235   \\
&&&&&&&&&& \\                                                                                                                                               
\hline                                                                                                                                                         
&&&&&&&&&& \\                                                                                                                                               
\end{tabular}
\end{center}
 
\parbox {7in} {A summary of the results on the hard, power-law
emission. Column 2 lists the measured photon indices ($\Gamma$).  The
normalizations of the power-law components ($A_1$) at an energy of
1keV are quoted in units of $10^{-5}$ photon
keV$^{-1}$cm$^{-2}$s$^{-1}$. Luminosities ($L_{\rm X}$) in the $1-10$
keV band are corrected for absorption and quoted in units of
$10^{40}$\ergps. The observed fluxes in the 2-10 keV band ($F_{\rm
X}$) are not corrected for absorption and are quoted in units of
$10^{-13}$ \ergpcmsqps. $\sigma_{\rm c}$ values are the confusion
limits for the S0 detectors in the $2-10$ keV band in units of
$10^{-13}$ \ergpcmsqps.  The $\Delta \chi^2$ values are the
improvements in $\chi^2$ obtained with the introduction of the
power-law component into the two-temperature models (either D or F as
appropriate). The total $\chi^2$ and degrees of freedom (DOF) for the
fits are also listed. The errors on $\Gamma$ and $A_1$ are the 90 per
cent confidence limits ($\Delta \chi^2 =2.71$) on a single interesting
parameter. The errors on the $L_{\rm X}$ and $F_{\rm X}$ values
account for the joint errors on the photon indices and
normalizations.}
\end{table*}

\begin{table}
\begin{center}
\caption{Additional constrains on the hard X-ray components using the 
bremsstrahlung model}
\vskip 0.1truein
\begin{tabular}{ c c c c c c c c c c c c }
\hline
         & ~ & $kT_{\rm brem}$   &  $A_2$       & ~ &  $\chi^2$/DOF \\
\hline
&&&&& \\   
M87      & ~ & $>9.8$    & $234^{+75}_{-41}$    & ~ &  1467.1/1143   \\
NGC 1399 & ~ & $>18.0$   & $33.0^{+5.7}_{-6.7}$ & ~ &  879.5/787   \\
NGC 4696 & ~ & $>29.0$   & $166^{+45}_{-40}$    & ~ &  956.0/746     \\
NGC 4472 & ~ & $>18.6$   & $23.7^{+4.2}_{-4.4}$ & ~ &  475.5/370    \\
NGC 4636 & ~ & $>8.4$    & $14.6^{+1.8}_{-1.1}$ & ~ &  473.1/227   \\
NGC 4649 & ~ & $>26.0$   & $21.2^{+4.7}_{-8.7}$ & ~ &  293.3/235   \\
&&&&& \\   
\hline
&&&&& \\   
\end{tabular}
\end{center}
\parbox {3.5in} {A summary of the results on the hard X-ray components 
using the bremsstrahlung model. Column 2 lists the measured electron 
temperatures ($kT_{\rm brem}$) in keV. Column 3 lists the normalizations 
($A_2$) in units of $10^{-5}$ photon keV$^{-1}$cm$^{-2}$s$^{-1}$. 
Column 4 summarizes the $\chi^2$ values (which are similar to those 
obtained with the power-law model in Table 4) and the numbers of 
degrees of freedom in the fits (DOF). Errors are 90 per cent 
confidence limits ($\Delta \chi^2=2.71$) on a single interesting parameter.} 
\end{table}

\subsubsection{Cosmic ray electrons and inverse Compton emission}

A third mechanism that could contribute to the detected X-ray fluxes
is inverse Compton emission due to primary cosmic ray electrons in
the intracluster medium. Detailed models (\eg Sarazin 1999) show that
for steady particle injection (with a given power-law
distribution) the inverse Compton spectra relax into a steady-state
form. Although these models predict ubiquitous EUV and soft X-ray
emission in clusters, due to electrons with $\gamma \sim 300$ (which
have the longest loss times and could be reasonably be injected since
$z \approxlt 1$), at harder energies their contribution is
negligible. Significant hard X-ray emission ($1-50 \keV$) can only be
produced by electrons with $\gamma \sim 10^3-10^4$ {\it i.e.} by
particles with rather short lifetimes (as set by inverse Compton
losses; $t_{\rm loss}\sim 10^9 \yr$) and would only be present in
clusters in which substantial particle injection has occurred since $z
<0.1$. Assuming that particles are accelerated in shocks in the
intracluster medium, one would only expect diffuse hard X-ray emission
in clusters undergoing, or having recently experienced, a major merger
event.  The Virgo cluster does not show evidence for recent merger
activity in its central regions (Allen \etal 1999). In
addition, even in cases were such particle injection does occur, the
expected steady state photon index where inverse Compton losses
dominate would be $\Gamma \approx 2.1$, significantly steeper than the
observed slopes of the power-law components.

Inverse Compton emission from the radio lobes is a further possible
source of power-law X-ray emission from the galaxies. Such emission is
normally very weak but has been detected from the hot-spots of
Cygnus-A (Harris, Carilli \& Perley 1994; Reynolds \& Fabian 1996) and
Fornax-A (Feigelson \etal 1995; Kaneda \etal 1995). The photon index
of the resulting X-ray emission should have a photon index similar to,
or steeper than, that of the radio emission which, for the radio
lobes, will be significantly steeper than $\Gamma \sim 1.2$. Thus,
although inverse Compton emission is expected at some level in all of
the systems, it is unlikely to contribute significantly to the hard,
X-ray components reported here.

\subsubsection{X-ray `reflection' from cold material}

\begin{table*}
\begin{center}
\caption{IRAS luminosities and 6.4 keV line limits}
\vskip 0.1truein
\begin{tabular}{ c c c c c c }
\hline       
         & ~ & $L_{\rm 60\mu m}$ & $L_{\rm 100\mu m}$ & $L_{\rm 100\mu
m}/L_{\rm X,1-10}$  &  E.W$_{\rm 6.4 keV}$  \\
\hline
&&&&& \\   
M87      & ~ & $78 \pm 9$   & $40 \pm 10$   & $0.95 \pm 0.36$  & $<90$eV    \\
NGC 1399 & ~ & $< 87$       & $81 \pm 26$   & $7.8 \pm 2.7$    & $<150$eV   \\
NGC 4696 & ~ & $300 \pm 90$ & $900 \pm 260$ & $4.2 \pm 2.1$    & $<190$eV   \\
NGC 4472 & ~ & $35 \pm 12$  & $21 \pm 17$   & $7.0 \pm 5.7$    & $<300$eV   \\
NGC 4636 & ~ & $31 \pm 7$   & $< 60$        & $< 22$           & $<130$eV   \\
NGC 4649 & ~ & $340 \pm 10$ & $1260 \pm 20$ & $490 \pm 60$     & $<210$eV   \\
&&&&& \\   
\hline                                                                     
&&&&& \\   
\end{tabular}
\end{center}

\parbox {7in} {A summary of the 60 and 100$\mu m$ IRAS luminosities for the
galaxies (in units of $10^{40}$\ergps) determined with the SCANPI
software and co-added IRAS scans. Error bars are the r.m.s. deviations of the 
residuals after baseline subtraction. Where no detection of a source was
made, an upper limit equivalent to 3 times the r.m.s. deviation is quoted. 
Column 4 lists the ratio of the 100$\mu m$ luminosities and $1-10$keV 
luminosities of the hard X-ray components. Column 5 lists the 90 per cent
confidence upper limits on the equivalent widths, E.W. (in eV, relative
to the power-law continua) of an intrinsic, narrow (instrumental 
resolution) 6.4 keV (Fe K) emission line in the ASCA spectra. }
\end{table*}

A final possibility for obtaining a relatively flat photon index in
the ASCA band is the situation where the observed X-ray flux is 
dominated by photons Compton scattered off cold,
optically-thick material close to the central X-ray source (\eg
Lightman \& White 1988; George \& Fabian 1991; Matt, Perola \& Piro
1991. This situation also requires that the primary X-ray source is
heavily obscured, a possibility argued against in Section 4.1). The
scattered X-ray spectrum will include absorption and emission features
due to various elements in the scattering medium and in particular
should exhibit a strong, fluorescent Fe-K emission line at 6.4
keV. ASCA observations of the Circinus galaxy (Matt \etal 1996) and
NGC 6552 (Reynolds \etal 1994) exhibit such `reflection-dominated'
spectra with strong (redshifted) 6.4 keV emission lines with
equivalent widths of $\sim 2$ keV. We have searched for the presence
of intrinsic 6.4keV emission lines in the ASCA spectra for the galaxies. 
In all cases we find no improvement to the fits 
obtained with spectral model G on the introduction of a narrow 6.4 keV 
emission line ($\Delta \chi^2=0.0$), and are able to place 90 per cent 
confidence limits ($\Delta \chi^2=2.71$) on the maximum equivalent widths 
(relative to the power-law continua) of any such lines
as listed in Table 6. We conclude that the flat, power-law 
components detected in the elliptical galaxies are unlikely to be due 
to this process.

\subsection{Summary of the results on the origin of the hard X-ray
components}

We have argued that the hard ($\Gamma \sim 1.2$) power-law, X-ray
components detected in the ASCA spectra are likely to be due to
bremsstrahlung emission from low-radiative efficiency accretion flows
onto the central supermassive black holes, coupled with
winds/outflows. This is a very different situation from Seyfert
galaxies, where the power-law X-ray emission is normally attributed 
to thermal inverse Compton scattering of the soft, disk radiation field.

We have examined a range of other mechanisms that could contribute
towards the detected X-ray fluxes. Some contribution towards the harder
X-ray emission could be made by the jets in these sources, although the
absence of a clear correlation between the 5 GHz radio and hard X-ray
luminosities suggests that the jet emission does not dominate the hard,
power-law flux. The close agreement between the ASCA and ROSAT (Section 5.1)
X-ray fluxes for M87, and the IRAS infrared measurements for the galaxies,
argue strongly against the observed flat photon indices and low luminosities
being simply due to intrinsic absorption acting on canonical Seyfert-like
spectra (due to accretion at the Bondi rates with a standard radiative
efficiency). Some level of intrinsic absorption is possible, however, and 
could explain
the very flat power-law components ($\Gamma < 1$) observed in some
objects (further flattening the spectra of the low-radiative efficiency
accretion flows, which should have an intrinsic  photon index in the ASCA 
band of $\Gamma \sim 1.4$). Limits on the 6.4 keV emission-line 
fluxes rule out a significant contribution to the detected hard, power-law 
components from X-rays Compton scattered off cold, optically-thick matter 
surrounding the central X-ray sources.

Binary X-ray sources are likely to contribute to the harder X-ray emission 
from the galaxies with lower-luminosity power-law components (the
Virgo ellipticals and NGC 1399). For M87 and NGC 4696, however, the
measured power-law luminosities are an order of magnitude greater than
the values predicted from the $L_{\rm X}/L_{\rm B}$ relation for
spiral and irregular galaxies. Moreover, the typical X-ray spectra of
X-ray binaries and black hole candidates are significantly steeper
than the observed hard, power-law components. Inverse Compton emission from
the radio lobes in the galaxies and primary cosmic-ray electrons in
the intracluster medium should not contribute significantly to the
detected power-law fluxes and, where present, will typically
produce a photon index in the ASCA band significantly steeper than the
measured values. Fluctuations in the Cosmic XRB should not
significantly effect our results.

Future X-ray observations at high spatial resolution made with the
Chandra Observatory will be crucial in establishing the contributions
from the various emission mechanisms outlined above and unambiguously
identifying the origin of the hard X-ray components. Deep X-ray
spectroscopy with XMM and ASTRO-E will allow us to examine variability in the
X-ray emission and search for broad, iron emission features associated
with the power-law components.  If the hard X-ray emission indeed
originates from bremsstrahlung processes in the outer regions of low
radiative-efficiency accretion flows, then the variability timescales
observed should be longer than those associated with typical Seyfert
nuclei. The detection of broad iron emission features would argue
against the simple, low-radiative efficiency accretion models
discussed here (Section 4.1), and require the presence of significant
amounts of cold, reflecting material close to the central black holes
(as in standard thin-disk accretion models).

\section{Previous constraints on nuclear X-ray emission from M87}

\subsection{ROSAT HRI observations} 

Reynolds \etal (1996) and Harris \etal (1997, 1998)
discuss ROSAT High Resolution Imager (HRI) observations of M87, and
measure a time-averaged HRI count rate associated with the active
nucleus of the source of $\sim 0.12$ count s$^{-1}$. We have used the
XSPEC software, incorporating the 1990 December version of the HRI
response matrix, to determine the flux of the power-law component
required to produce the count rate observed in the $0.1-2.4$ keV ROSAT
HRI band. For an assumed power-law emission model with a photon index,
$\Gamma=1.4$, absorbed by a Galactic column density, $N_{\rm H} = 2.5
\times 10^{20}$ \apc, we find that a normalization, $A_1 \sim 1.5
\times 10^{-3}$ photon keV$^{-1}$cm$^{-2}$s$^{-1}$ is required to
account for the observed HRI count rate.  This is in excellent
agreement with the value measured from the ASCA spectra of $A_1=
1.4^{+1.6}_{-0.6} \times 10^{-3}$ photon keV$^{-1}$cm$^{-2}$s$^{-1}$.
(We note that accounting for the intrinsic absorption also detected in
the ASCA spectra would imply a larger intrinsic normalization for the
power-law component of $A_1 \sim 5.0 \times 10^{-3}$ photon
keV$^{-1}$cm$^{-2}$s$^{-1}$.  However, this corrected value assumes
that the absorption is due to cold gas in a uniform screen in front of
the nucleus, which may not be the case \eg Allen \& Fabian 1997;
Reynolds 1997).

\subsection{Previous analyses of the ASCA observations} 

Two previous studies of M87 have also reported results based on the
same ASCA observations discussed in this paper. Matsumoto \etal (1996)
present results from an analysis using a simple two-temperature plasma
plus power-law spectral model, with a fixed power-law photon index of
$\Gamma = 1.7$, from which they determine a flux associated with the
nucleus (in the $0.5-4.5$ keV band) of $1.1 \pm 0.2 \times
10^{-11}$\ergpcmsqps. This flux is $\sim 2.5$ times larger than the
value implied by our best-fitting spectral model (model G).  Reynolds
\etal (1996) also present results from a study of the same ASCA
observations, in which they did not detect significant emission 
associated with the nucleus, although these authors did not account 
for the possibility of variable element abundance ratios in their 
analysis, which provides a crucial step in the modeling (see also Allen 
\etal 1999).  To further illustrate this, we have repeated 
our analysis of the power-law component in M87 with the element
abundances linked to vary in the same ratio relative to their solar
values ({\it c.f.} spectral model D). The best-fit $\chi^2$ value
obtained with this more simple model, $\chi^2=1888$, is substantially
worse than the value, $\chi^2=1468$, obtained with spectral model
G. The best-fit parameter values ($\Gamma = 1.70^{+0.27}_{-0.31}$ and
$A_1=1.9^{+1.5}_{-0.9} \times 10^{-3}$ photon
keV$^{-1}$cm$^{-2}$s$^{-1}$) are also (slightly) offset from the
results obtained with model G. Such differences demonstrate the need
to fully account for the complex temperature structure of the
galaxy/cluster plasma and possible variations in individual element
abundances ratios when attempting to constrain the power-law emission
from such sources.

\subsection{Rossi X-ray Timing Explorer observations} 

Reynolds \etal (1999) present further constraints on power-law
emission from M87 using observations made with the Proportional
Counter Array (PCA) on the Rossi X-ray Timing Explorer (RXTE). The
data presented by these authors cover the $3-15$ keV range and
constrain the $2-10$ keV flux of the power-law component to be $< 4.1
\times 10^{-12}$ \ergpcmsqps, for an assumed photon index $\Gamma
=2.0$. This upper limit is lower than the measured value of
$8.7^{+1.7}_{-1.6} \times 10^{-12}$ \ergpcmsqps~from the ASCA data
(Table 4). The ASCA data also require a significantly flatter photon
index than assumed in the RXTE study.

The best-fitting spectral model for M87 determined from the ASCA
analysis and plotted in Fig. 2 shows that the extended cluster
emission dominates over the power-law component in the ASCA spectra
across virtually the entire energy range of the detectors. Only at
energies $E > 8-9$ keV does the power-law component dominate the
detected flux. The much larger field of view of the PCA collimator
($\sim 1$ degree$^2$ FWHM; Jahoda \etal 1996) results in a larger
fraction of the total cluster flux ($\sim 5$ times more) being
included in the detected spectrum (the cluster emission extends to a
radius of at least 4 degree; Schindler, Binggeli \& B\"ohringer
1999). Thus, the power-law component, as determined from the ASCA
data, will not dominate over the cluster emission in the PCA spectrum
below an energy of $\sim 12-13$ keV.

Modelling the extended plasma emission in a cluster as cool as Virgo Cluster
(Table 3) with instruments like the PCA, restricted to the (relatively hard) 
$3-15$keV energy range, is difficult. The PCA spectra cannot reliably 
constrain the multiphase nature of the gas in the cluster core which, 
as this study has shown, can be crucial in constraining the properties of 
the power-law emission. However, although such considerations may be relevant 
in interpreting the RXTE results, it remains plausible that the nuclear 
emission from M87 may simply have varied between the 
ASCA observations in 1993 June and the PCA observations made in 1998 
January (Harris \etal 1997, 1998; Tsvetanov \etal 1998)

\subsection{ROSAT HRI flux-limits on other sources in the sample}

The other galaxies discussed in this paper have not previously been
studied in the same detail as M87 and no detections of point-source
X-ray emission associated with their nuclei have been reported.
However, Di Matteo \etal (1999a) present limits on possible nuclear
X-ray emission for the three Virgo ellipticals, based on ROSAT HRI
imaging data.  Their limits, which are defined at an energy of 1keV,
are $\nu F_\nu < 6.8 \times 10^{-14}$ \ergpcmsqps (NGC 4472), $\nu
F_\nu < 7.8 \times 10^{-14}$ \ergpcmsqps (NGC 4636) and $\nu F_\nu <
1.5 \times 10^{-13}$ \ergpcmsqps (NGC 4649).  These limits compare to
the measured ASCA fluxes, quoted in the same units, of $\nu F_\nu \sim
6.1 \times 10^{-14}$ \ergpcmsqps (NGC 4472), $\nu F_\nu \sim 1.6
\times 10^{-13}$ \ergpcmsqps (NGC 4636), and $\nu F_\nu \sim 3.5
\times 10^{-14}$ \ergpcmsqps (NGC 4649).

The ASCA measurements for NGC 4472 and 4649 are within the Di Matteo
\etal (1999a) limits. For NGC 4636, however, the ASCA measurement is
approximately twice the ROSAT limit. The Di Matteo \etal (1999a)
limits are determined by fitting an analytic King model to the
observed X-ray surface brightness profiles and determining the maximum
additional contribution that can be made by a central point
source. However, these limits are sensitive to complexities in the
observed surface brightness profiles and, especially for NGC 4636,
which exhibits a complex X-ray morphology in its central regions, the
ROSAT limits should be viewed with caution.

\section{The element abundances in NGC 4636}

In contrast to the ASCA results for M87 and NGC 4696 (Allen \etal
1999), for which the introduction of individual element abundances as
free parameters in the fits leads to a more significant improvement in
the statistical quality of the fits than the introduction of the
power-law component, for NGC 4696, the-introduction of the power-law
emission component provides by far the most significant improvement
over the basic two-temperature model.  Thus, in determining our
results on the abundances of the individual elements in NGC 4636, we
started from the two-temperature model with the power-law component
included, which provides a $\chi^2$ of 722.6 for 230 degrees of
freedom. We then systematically determined the
statistical improvements to the fit obtained by allowing the abundance
of each element, in turn, to be a free parameter in the analysis.
Having identified the element providing the most significant
improvement, the abundance of that element was maintained as a free
parameter, and the process repeated to determine the element providing
the next most significant improvement.

In agreement with the results for M87 and NGC 4696, we find that the 
most significant improvements in the fit to the NGC 4636 data 
are obtained by including the abundances of Si, Mg and S
as free fit parameters (the measured $\chi^2$ value is reduced from
722.6 to 475.0, for the introduction of only three extra fit
parameters).  At this point, including the abundances of further
elements as fit parameters did not lead to such significant
improvements and, due to the already complex nature of the best-fit
model, we forced the abundances of all other elements to vary with 
the same ratio relative to their solar values (effectively tying the
abundances to that of iron, the element to which the ASCA data are most
sensitive). We note, however, that including the abundances of
Na and O as further free fit parameters did lead to formally
significant improvements, with reductions in $\chi^2$ of $ \sim 20$
and 30, respectively. However, due to the systematic uncertainties in
the NGC 4636 spectra at low energies (Section 3.3), which effect the 
measured O abundance, and the fact that the Na abundance fits to an 
un-physically high value ($\sim 5$ times solar), the metallicities 
of these elements were not included as free parameters in our final 
analysis). The measured element abundances for NGC 4636, as a fraction of their
solar photospheric values defined by Anders \& Grevesse (1989), are
$Z_{\rm Fe} = 0.62^{+0.13}_{-0.09}$, $Z_{\rm Mg} =
0.93^{+0.19}_{-0.15}$, $Z_{\rm Si} = 0.96^{+0.18}_{-0.12}$ and $Z_{\rm
S} = 1.49^{+0.27}_{-0.22}$. These results are in reasonable agreement with
those of Matsushita \etal (1997) and Buote (1999).

Scaling our measured abundance ratios to the meteoric abundance scale
of Anders \& Grevesse (1989) we determine [Mg/Fe] $\sim 0.00$, [Si/Fe]
$\sim 0.02$ and [S/Fe]$\sim 0.16$. Comparing these values with the
supernova yield models of Nagataki \& Sato (1998), our observed
[Si/Fe] ratio implies a mass fraction of the iron enrichment due to
type Ia supernova, $M_{\rm Fe,SNIa}/M_{\rm Fe,total}$, in the range
$0.55-0.9$ (where the limits cover the full range of models examined
by these authors).  For spherical type II supernovae, a mass fraction
in the range $M_{\rm Fe,SNIa}/M_{\rm Fe,total} \sim 0.7-0.85$ is
preferred. Further comparison of our [Si/Fe] constraint with the
supernovae models discussed by Gibson \etal (1997) also requires
$M_{\rm Fe,SNIa} \sim 0.5-0.7$.  However, the observed [S/Fe] and
[Mg/Fe] ratios favour a mass fraction due to type Ia supernovae of
$\approxlt 0.5$ (Gibson \etal 1997).

\section{Conclusions}

We have presented detections of hard X-ray emission components in the 
spectra of six, nearby giant elliptical galaxies observed with ASCA. The 
galaxies exhibit clear, dynamical evidence for supermassive ($10^8-$ a few 
$10^9$\Msun) black holes in their nuclei. The hard X-ray emission can be
parameterized by power-law models with photon indices, $\Gamma =
0.6-1.5$ (mean value 1.2), and luminosities, $L_{\rm X,1-10} \sim 2.6
\times 10^{40}-2.1 \times 10^{42}$ \ergps, or thermal bremsstrahlung 
models with electron temperatures, $kT > 10$ keV. Such properties 
identify these galaxies as a new class of accreting X-ray source, with 
X-ray spectra significantly harder than those of Seyfert nuclei, typical 
binary X-ray sources and low-luminosity AGN, and bolometric luminosities 
comparatively dominated by their X-ray emission. We have argued that the 
hard X-ray emission is likely to be due to accretion onto the central, 
supermassive black holes, via low-radiative efficiency flows, coupled 
with strong outflows. Within such models, the hard X-ray emission originates 
from bremsstrahlung processes in the radiatively-dominant, outer regions of 
the accretion flows. (Detailed modeling and discussion of these issues are 
presented by Di Matteo \etal 1999b).

For the case of M87, the flux of the hard component was shown to be in
good agreement with the nuclear X-ray flux determined from earlier
ROSAT HRI observations, which were able to resolve knot A in the jet
from the nuclear emission component. We have stressed the importance
of accounting for the complex temperature structure, intrinsic
absorption and variable element abundance ratios in the analysis of
the ASCA spectra. We confirmed results showing that the application of
such models leads to measurements of approximately solar emission-weighted 
metallicities for the X-ray gas in the galaxies. We also presented detailed 
results on the individual element abundances in NGC 4636.

Future observations at high spatial resolution with the Chandra
Observatory will be crucial in establishing the contributions from the
various X-ray emission mechanisms present in elliptical galaxies and
unambiguously identifying the origin of the hard X-ray components.  
Deep X-ray spectroscopy with XMM and ASTRO-E will allow us to 
examine variability in the X-ray emission (which should be slower in sources
where the X-ray emission originates from the outer regions of low
radiative-efficiency accretion flows than in typical Seyfert nuclei)
and to search for broad, iron emission features associated with the
power-law components. The detection of such broad emission features
would argue against the simple, low-radiative efficiency accretion
models discussed here, and require the presence of significant amounts
of cold, reflecting material close to the central black holes.

The discovery of hard X-ray emission components in the spectra of nearby elliptical 
galaxies containing supermassive black holes provides important new 
constraints on the accretion processes in these systems. Our results are 
relevant to understanding the demise of quasars (which could plausibly be due 
to a change in the dominant accretion mode in ellipticals over the history of 
the Universe) and, ultimately, the origin of the hard ($\Gamma \sim 1.4$) 
Cosmic X-ray Background (\eg Di Matteo \& Fabian 1997b). These issues, and 
others, will be explored in future papers 
(Di Matteo \etal 1999b; Di Matteo \& Allen 1999).

\section*{Acknowledgements}
We thank Ramesh Narayan, Eliot Quataert and Chris Reynolds for useful 
discussions. TDM acknowledges support for this work provided by NASA 
through AXAF Fellowship grant number PF8-10005 awarded by the AXAF 
Science Center, which is operated by the Smithsonian Astrophysical 
Observatory for NASA under contract NAS8-39073.


\clearpage

\clearpage


\end{document}